\documentclass[aps,prd,10pt,superscriptaddress,nofootinbib]{revtex4-2}

\usepackage[utf8]{inputenc}
\usepackage[T1]{fontenc}
\usepackage{anyfontsize}
\usepackage{amsmath}
\usepackage{amssymb}
\usepackage{amsfonts}
\usepackage{mathrsfs}
\usepackage{enumitem}

\usepackage{microtype}
\usepackage[normalem]{ulem}

\usepackage{graphicx}
\usepackage{epsfig}
\usepackage{float}

\usepackage[dvipsnames]{xcolor}
\usepackage{hyperref}
\hypersetup{
    colorlinks=true,
    citecolor=Purple,
    linkcolor=Purple,
    urlcolor=Purple,
    linktocpage=true,
    breaklinks=true
}
\usepackage{soul}
\usepackage[capitalize]{cleveref}

\begin{document}

\title{Traversable Hyperbolic Wormholes with a Casimir-Memory Source: Complexity and Shell-Free Matching}

\author{Celio R. Muniz}
\email{celio.muniz@uece.br}
\affiliation{Universidade Estadual do Cear\'a (UECE), Faculdade de Educa\c{c}\~ao, Ci\^encias e Letras de Iguatu, Av. D\'ario Rabelo s/n, Iguatu - CE, 63.500-00 - Brasil}
\author{Roberto Avalos}
\email{roberto.avalos@emory.edu}
\affiliation{Department of Physics, Emory University, Atlanta, GA 30322, USA}
\author{Jonathan A. Rebouças}
\email{jalvesreboucas@ifce.edu.br}
\affiliation{Instituto Federal de Educação Ciências e Tecnologia do Ceará (IFCE), Iguatu-CE, Brasil}
\author{Francisco Bento Lustosa}
\email{chico.lustosa@uece.br}
\affiliation{Universidade Estadual do Cear\'a (UECE), Faculdade de Educa\c{c}\~ao, Ci\^encias e Letras de Iguatu, Av. D\'ario Rabelo s/n, Iguatu - CE, 63.500-00 - Brasil}
\author{Francisco Tiago Barboza Sampaio}
\email{tiago.barboza@uece.br} 
\affiliation{Universidade Estadual do Cear\'a, Faculdade de Educa\c c\~ao, Ci\^encias e Letras de Iguatu, 63500-000, Iguatu, CE, Brasil.}

\begin{abstract}
Hyperbolically symmetric wormholes provide a natural setting in which negative energy densities arise directly from the throat geometry and finite matter configurations can be smoothly joined to the hyperbolic Schwarzschild vacuum. We construct such configurations from a matter-first perspective, using an effective Casimir source corrected by a gravitational-memory contribution, $\rho(r)=-\alpha/r^4+\eta/r^7$. The $r^{-7}$ term is motivated by the persistent shift of the Casimir vacuum energy produced by a transient gravitational perturbation, while its promotion to a radial source is treated phenomenologically. The density determines the shape function analytically, and the temporal geometry is then fixed through complexity-based conditions. Full vanishing complexity, $Y_{TF}=0$, is first imposed and subsequently relaxed through a one-parameter deformation with residual complexity $Y_{TF}=(p-3)r_0/(2r^3)$. Decomposing the complexity factor into local and throat contributions shows that local vanishing complexity is not an independent closure, but occurs at the distinguished value $p=\beta'(r_0)$. At this value, besides the constant-redshift solution of the finite-$r$-derivative branch, a second lapse branch is regular in proper radial distance and admits a semianalytical representation. We further show analytically that pressure-free Darmois matching in the finite-$r$-derivative family requires $p<\beta'(r_0)$, thereby enforcing tangential null and strong-energy-condition violations at the throat throughout the matching-admissible domain. The common density remains negative throughout the matter-supported region, implying weak- and dominant-energy-condition violations independently of the lapse branch. The memory parameter modifies the flare-out domain, redshift profile, and matching data, while finite pressure-free junctions permit smooth matching to the hyperbolic Schwarzschild vacuum without thin shells. The construction thus combines a memory-corrected Casimir source with a unified complexity framework in which the different temporal sectors emerge as systematically related branches of the same radial geometry.
\end{abstract}


\maketitle

\section{Introduction}
\label{introdução}

Traversable wormholes are among the most striking geometrical solutions allowed by General Relativity. Since the seminal work of Morris and Thorne, they have been studied as spacetime configurations in which two regions are connected through a throat that can, in principle, be crossed without encountering an event horizon. Their main physical difficulty is equally well known: in Einstein gravity, the flaring-out of the throat is generically tied to violations of the null energy condition (NEC), which motivates the search for quantum or effective sources capable of supplying negative energy densities in a controlled way \cite{Morris:1988cz,Visser:1995cc,Lobo:2017cay}. Quantum inequalities further constrain the magnitude and duration of negative-energy configurations, making the nature, localization, and origin of the required energy-condition violation central issues in wormhole physics \cite{Ford:1995wg,PhysRevLett.90.201102}.

Among the few experimentally grounded manifestations of negative vacuum energy, the Casimir effect occupies a particularly important place. Originally predicted for conducting boundaries as a consequence of quantum vacuum fluctuations \cite{Casimir:1948dh}, and subsequently measured with high precision \cite{PhysRevLett.78.5}, it provides a concrete example of the dependence of the quantum vacuum on boundary conditions \cite{BORDAG20011}. This has motivated a broad literature on Casimir-supported wormholes, both in Einstein gravity and in modified-gravity settings \cite{Garattini:2019ivd,Santos:2023zrj,Muniz:2024jzg,Muniz:2025teu,Pourhassan:2025htk,Sarkar:2025iiz,Alblowy:2026hfr}. In such constructions the Casimir sector is not merely a formal source of exotic stress energy: it supplies a semiclassically motivated mechanism for generating the negative vacuum contribution required by the geometry.

Most wormhole models are formulated with spherical symmetry. Hyperbolically symmetric configurations provide a qualitatively different setting, however, because the sign structure of the Einstein equations changes the relation between the throat geometry and the matter variables. Hyperbolic wormholes and related configurations have therefore attracted increasing attention \cite{Mimoso:2010yp,Cataldo:2015vra,Zubair:2022jjm,Choudhury:2025ysx}. In parallel, Herrera and collaborators have developed a description in which static hyperbolically symmetric fluids and vacuum geometries are naturally associated with the region inside the Schwarzschild horizon \cite{Herrera:2018mzq,Herrera:2020bfy,Herrera:2025jsu}. Within this interpretation, a finite hyperbolic matter distribution may be matched to the corresponding vacuum sector at a radius below $2M$, rather than being required to extend to an asymptotically flat region.

A particularly relevant antecedent is the hyperbolic Casimir-like wormhole constructed by Avalos \emph{et al.} \cite{Avalos:2025hfw}. That work showed that a traversable hyperbolic wormhole can be smoothly joined to the hyperbolic vacuum through the Darmois conditions, without introducing a thin shell, and used the complexity factor as one of the supplementary ingredients required to close the Einstein system. Their construction starts from a prescribed metric/complexity sector and obtains a Casimir-like matter configuration compatible with it. The present work adopts the complementary, matter-first viewpoint: we prescribe the vacuum-energy profile, determine the shape function directly from the density equation, and only then use the complexity sector to determine the redshift function. This reversal is important because it allows the geometry to retain explicit information about the physical parameters entering the chosen quantum-vacuum source.

The second ingredient of our construction is gravitational memory. Gravitational-wave memory describes a residual effect that survives after a transient gravitational disturbance has passed \cite{Braginsky:1987kwo,PhysRevLett.67.1486}; its modern formulation is also closely connected with asymptotic symmetries and soft-graviton physics \cite{Strominger:2014pwa,Pasterski:2015tva}. Of particular relevance here, Sorge \cite{Sorge:2023sdf} studied a massless quantum field confined to a Casimir cavity in a weak, time-dependent anisotropic background and found a permanent shift of the zero-point energy after the perturbation has disappeared. For a weak gravitational-wave pulse, this residual correction depends on the parameters of the pulse and falls with a higher power of the cavity size than the ordinary Casimir term. Thus, the Casimir vacuum can retain a small imprint of the previous gravitational dynamics even when the spacetime has returned to its asymptotic state.

A closely related implementation of this idea was recently explored in the
spherically symmetric Morris--Thorne framework, where the same
Casimir-memory-inspired density profile was employed to construct
traversable wormholes \cite{Reboucas:2026memory}. In that setting, the
redshift sector was closed through a constant barotropic equation of state
together with throat regularity. The present work addresses a different
geometric and structural problem: hyperbolic symmetry changes the relation
between the throat geometry and the matter variables, while the remaining
metric freedom is determined through complexity-based closure conditions
rather than through a prescribed equation of state. Thus, while the two
constructions share the same effective vacuum source, they probe distinct
symmetry sectors and closure mechanisms.

Motivated by this distinction, we investigate how a hyperbolic Casimir
wormhole is modified when the vacuum source carries a residual
gravitational-memory contribution. We adopt the phenomenological effective
density
\begin{equation}
\rho(r)=-\frac{\alpha}{r^4}+\frac{\eta}{r^7},
\end{equation}
where the first term represents the ordinary Casimir-like contribution and
the second parametrizes the memory-inspired correction. The coefficient
$\eta$ is related to the pulse-dependent memory amplitude in Sorge's
calculation; its detailed identification, including its dependence on the
gravitational-wave amplitude and time scale, is given in Sec.~III. We
emphasize that Sorge's result is derived for a Casimir cavity whose physical
scale is the plate separation. The replacement of that scale by the radial
variable, producing the effective $r^{-7}$ profile above, is a
phenomenological modeling step of the present work rather than a claim that
a hyperbolic $r^{-7}$ density was derived in Ref.~\cite{Sorge:2023sdf}.

Our aim is not merely to add this term to an existing wormhole solution.
Once the density is prescribed, the Einstein equation fixes the shape
function $\beta(r)$ analytically, so the Casimir-memory parameters enter
the radial geometry from the outset. The remaining metric freedom is then
investigated through the complexity factor $Y_{TF}$, a geometrically
defined structure scalar arising from the orthogonal splitting of the
Riemann tensor. It encodes the combined effects of pressure anisotropy
and energy-density inhomogeneity and provides a well-established framework
for characterizing the structural complexity of self-gravitating systems
\cite{Herrera:2018complexity}. In the wormhole context, the complexity
factor has also been employed as a supplementary closure prescription for
the Einstein equations, including in Casimir-supported configurations
\cite{Avalos:2022complexity,Bhattacharya:2023complexity}.

We first impose exact vanishing complexity, $Y_{TF}=0$, and construct the
corresponding redshift function subject to throat regularity and smooth
Darmois matching. We then relax this condition through a one-parameter
deformation. As shown below, the deformation has a direct interpretation,
\begin{equation}
Y_{TF}(r)=\frac{(p-3)r_0}{2r^3},
\end{equation}
so that $p-3$ measures the amplitude of a controlled residual complexity
rather than acting as an arbitrary numerical parameter. A further
decomposition of $Y_{TF}$ into local and throat contributions then reveals
an additional structural relation,
\begin{equation}
Y_{TF}^{\rm loc}=0
\qquad\Longleftrightarrow\qquad
p=\beta'(r_0).
\end{equation}
Thus local vanishing complexity is not an independent third closure: it
occurs at a distinguished value of the deformation parameter. At this
value, the finite-$r$-derivative solution contains a constant-redshift
branch, while a second independent branch yields a nonconstant lapse that
is regular when expressed in terms of proper radial distance.

The resulting framework therefore differs from earlier hyperbolic Casimir constructions in three connected respects. First, the matter sector is fixed before the redshift function, so that the Casimir-memory source determines the shape function directly. Second, the vacuum source contains a memory-inspired contribution tied to the parameters of a preceding gravitational pulse. Third, instead of employing a single complexity prescription, we organize the temporal sector through an exact vanishing-complexity solution and its controlled one-parameter deformation, and then identify local vanishing complexity as the distinguished sector $p=\beta'(r_0)$ of the same deformation. The latter sector contains two lapse branches: a constant-redshift finite-$r$-derivative solution and a second nonconstant branch that is regular in proper radial distance. For the nonconstant branch we derive the general Darmois matching while retaining the relative time normalization between the interior and vacuum regions; the normalized choice $\lambda=1$ then appears as a particular subclass, whose admissible domain is studied separately. In this way the different temporal sectors are not independent wormhole models, but systematically related branches supported by the same Casimir-memory radial geometry.

A further objective is to determine which properties are genuinely associated with the memory correction and which follow already from hyperbolic geometry. In particular, the hyperbolic flaring-out condition itself enforces negative energy density and radial NEC violation at the throat in Einstein gravity. We therefore examine the energy conditions throughout the finite matter-supported region, rather than identifying exoticity solely with the sign of the radial pressure. We also determine the parameter domains for throat flaring-out, regularity, Euclidean embeddability when that representation is required, and smooth matching to the hyperbolic Schwarzschild vacuum. The resulting configurations are finite interior solutions with $r_\Sigma<2M$, in the same global interpretation adopted for hyperbolic Schwarzschild interiors \cite{Herrera:2018mzq,Herrera:2020bfy,Herrera:2025jsu,Avalos:2025hfw}; they are not independently asymptotically flat wormholes.

The paper is organized as follows. Section~II introduces the hyperbolic wormhole geometry, the Einstein equations, throat conditions, and the Euclidean-embedding criterion. In Sec.~III we specify the Casimir-memory density, relate the memory parameter to Sorge's result, and derive the shape function and its admissible throat domain. Section~IV develops the exact and $p$-deformed complexity sectors, their regularity conditions, and their smooth matching to the hyperbolic vacuum, and establishes the relation $Y_{TF}^{\rm loc}=0\Leftrightarrow p=\beta'(r_0)$. Section~V analyzes the second, nonconstant lapse branch at this distinguished value, including its proper-distance regularity, general Darmois matching, and normalized $\lambda=1$ subclass. Section~VI analyzes the energy conditions, separating the source- and throat-level statements common to all lapse branches from the pressure-dependent results specific to the finite-$r$-derivative $p$-family. Finally, Sec.~VII summarizes the results and their physical implications.

\section{Hyperbolic Wormhole Geometry and Field Equations}
\label{sec:geometry}

We begin by considering a static spacetime endowed with hyperbolic symmetry. Unlike the standard Morris--Thorne wormhole, whose angular sector is foliated by two-dimensional spheres of positive curvature, the present construction is based on two-dimensional hyperbolic surfaces of constant negative curvature. Such geometries arise naturally in several gravitational contexts and possess distinctive properties regarding the behavior of matter sources, energy conditions, and matching procedures. In particular, hyperbolic symmetry provides a natural framework for accommodating negative energy densities, making it especially suitable for the construction of exotic compact objects sustained by vacuum-like sources.

The spacetime line element is assumed to be
\begin{equation}
ds^2=
-e^{2\Phi(r)}dt^2
+\frac{dr^2}{\frac{\beta(r)}{r}-1}
+r^2d\theta^2
+r^2\sinh^2\theta\,d\phi^2,
\label{eq:metric}
\end{equation}
where $\Phi(r)$ is the redshift function and $\beta(r)$ denotes the shape function. The latter determines the spatial geometry of the wormhole and controls the existence of the throat. As in conventional wormhole geometries, the throat is located at a radius $r=r_0$ satisfying
\begin{equation}
\beta(r_0)=r_0.
\end{equation}
However, due to the negative curvature of the angular sector, the corresponding flaring-out condition differs from that of spherically symmetric wormholes. In the hyperbolic case, the flaring-out condition requires
\begin{equation}
\beta'(r_0)>1,
\end{equation}
which guarantees that the spatial section opens outward at the throat. This condition is the counterpart of the familiar Morris--Thorne requirement $\beta'(r_0)<1$, highlighting one of the fundamental geometric differences between spherical and hyperbolic wormholes.

It is also useful to distinguish the local flaring-out requirement from a global condition associated with the usual Euclidean embedding diagram. Consider a constant-time spatial section and choose a constant hyperbolic angle $\theta=\theta_*$ such that $\sinh\theta_*=1$. The induced two-dimensional metric is
\begin{equation}
 ds^2=\frac{dr^2}{\beta(r)/r-1}+r^2d\phi^2.
 \label{eq:embedding_slice}
\end{equation}
Embedding this surface in Euclidean three-space with cylindrical line element $ds_E^2=dr^2+r^2d\phi^2+dz^2$ gives
\begin{equation}
\left(\frac{dz}{dr}\right)^2=\frac{2r-\beta(r)}{\beta(r)-r}.
\label{eq:embedding_derivative}
\end{equation}
Hence a real Euclidean embedding requires, away from the throat,
\begin{equation}
 r<\beta(r)\leq 2r,
 \label{eq:embedding_condition}
\end{equation}
with $\beta(r_0)=r_0$ at the throat. The lower inequality is the static hyperbolic-domain condition, whereas the upper inequality ensures that the particular Euclidean embedding used here remains real. We emphasize that $\beta(r)\leq2r$ is therefore an embeddability condition for this spatial representation, rather than an independent requirement for the existence of the Lorentzian metric itself.

The matter source is modeled by an anisotropic fluid with energy--momentum tensor
\begin{equation}
T^\mu_{\ \nu}
=
\mathrm{diag}
\left(
-\rho,
p_r,
p_\perp,
p_\perp
\right),
\label{eq:stress_tensor}
\end{equation}
where $\rho(r)$, $p_r(r)$, and $p_\perp(r)$ denote the energy density, radial pressure, and tangential pressure, respectively. Anisotropic stresses arise naturally in hyperbolically symmetric configurations and are generally unavoidable in the presence of exotic matter sources.

Substituting the metric \eqref{eq:metric} into Einstein's field equations,
\begin{equation}
G_{\mu\nu}=8\pi T_{\mu\nu},
\end{equation}
one obtains the gravitational field equations for hyperbolic symmetry \cite{Herrera:2021symmetry}
\begin{align}
8\pi\rho
&=
-\frac{\beta'}{r^2},
\label{eq:rho_einstein}
\\
8\pi p_r
&=
\frac{\beta}{r^3}
+
\frac{2\Phi'}{r}
\left(
\frac{\beta}{r}-1
\right),
\label{eq:pr_einstein}
\\
8\pi p_\perp
&=
\left(
\frac{\beta}{r}-1
\right)
\left(
\Phi''+\Phi'^2
\right)
+
\frac{\Phi'}{2r^2}
\left(
r\beta'+\beta-2r
\right)
+
\frac{r\beta'-\beta}{2r^3}.
\label{eq:pperp_einstein}
\end{align}
Although these equations resemble their spherical counterparts, the negative curvature of the hyperbolic angular sector introduces important sign changes that substantially modify the relationship between geometry and matter. As a consequence, several qualitative features of hyperbolic wormholes differ from those of standard Morris--Thorne configurations.

A particularly important result follows directly from Eq.~\eqref{eq:rho_einstein}. Since
\begin{equation}
8\pi\rho(r_0)
=
-\frac{\beta'(r_0)}{r_0^2},
\label{eq:throat_density}
\end{equation}
the flaring-out condition immediately implies
\begin{equation}
\rho(r_0)<0.
\end{equation}
Thus, within Einstein gravity, a negative energy density at the throat follows directly from the hyperbolic flaring-out geometry, independently of the particular density profile adopted below.

An equally important consequence concerns the null energy condition. Evaluating the combination $\rho+p_r$ at the throat yields
\begin{equation}
8\pi(\rho+p_r)\Big|_{r_0}
=
\frac{1-\beta'(r_0)}{r_0^2}.
\label{eq:throat_radial_nec}
\end{equation}
Since $\beta'(r_0)>1$, it follows that
\begin{equation}
\rho+p_r<0,
\end{equation}
demonstrating that the null energy condition is necessarily violated at the throat. Thus, in hyperbolic wormholes, the exotic nature of the matter source is not merely a consequence of the chosen energy density profile but is directly tied to the geometric requirements imposed by traversability.

The system described by Eqs.~\eqref{eq:rho_einstein}--\eqref{eq:pperp_einstein} contains five unknown functions, namely $\rho(r)$, $p_r(r)$, $p_\perp(r)$, $\Phi(r)$, and $\beta(r)$, while Einstein's equations provide only three independent relations. Two additional inputs are therefore required to close the system. In the present construction, these are supplied by prescribing the energy-density profile and by imposing a condition on the complexity factor.

In the present work, we proceed differently from the hyperbolic Casimir wormhole construction of Ref.~\cite{Avalos:2025hfw}. Rather than introducing an equation of state relating the thermodynamic variables, we specify a physically motivated energy density inspired by Casimir vacuum fluctuations modified by gravitational-wave memory effects. The shape function is then obtained directly from Einstein's equations, while the remaining freedom associated with the redshift function is fixed through complexity-based closure conditions, beginning with exact vanishing complexity. This approach allows the spacetime geometry to emerge from the combined interplay of the matter distribution and the chosen closure prescription.

In the next section, we introduce the Casimir-memory energy density profile
and derive the corresponding shape function analytically. The resulting
geometry will provide the common radial background for the complexity
sectors developed in the following sections.

\section{Shape Function from the Casimir--Memory Density}
\label{sec:shape}

Having established the gravitational field equations, we now specify the matter source supporting the wormhole geometry. Motivated by Sorge's analysis of the gravitational memory of the Casimir effect~\cite{Sorge:2023sdf}, we adopt the phenomenological effective density profile
\begin{equation}
\rho(r)=-\frac{\alpha}{r^4}+\frac{\eta}{r^7},
\qquad
\alpha>0,\quad \eta\geq 0.
\label{eq:rho}
\end{equation}
Here $\alpha$ characterizes the ordinary Casimir sector, while $\eta$ parametrizes a memory-inspired positive correction. More specifically, Sorge showed that, for a weak Gaussian gravitational pulse acting on a Casimir cavity, the late-time vacuum energy of a massless scalar field acquires the leading residual contribution~\cite{Sorge:2023sdf}
\begin{equation}
\delta\rho_{\rm mem}(L)
\simeq
\frac{15H^2}{64\sqrt{2}\,\pi\,\sigma^3 L^7},
\label{eq:sorge_memory}
\end{equation}
where $H$ is the strain amplitude of the gravitational perturbation and $\sigma^{-1}$ characterizes its duration. For an electromagnetic field the correction is doubled because of the two polarization states. Accordingly, for the scalar-field result one may identify
\begin{equation}
\eta_{\rm S}=\frac{15H^2}{64\sqrt{2}\,\pi\,\sigma^3},
\label{eq:eta_sorge}
\end{equation}
with twice this value for the electromagnetic case. In Sorge's calculation $L$ is the plate separation. Our wormhole source is obtained by promoting this separation scale to an effective radial scale, $L\rightarrow r$. Thus, the $r^{-7}$ term in Eq.~\eqref{eq:rho} should be understood as a phenomenological radial implementation of Sorge's late-time Casimir-memory scaling, rather than as a spherically or hyperbolically symmetric density profile derived directly from the gravitational-wave calculation. In the analysis below we consequently retain $\eta$ as an effective nonnegative parameter, with Eq.~\eqref{eq:eta_sorge} providing its microscopic motivation and its dependence on the pulse amplitude and duration in the weak-pulse regime. Substituting Eq.~\eqref{eq:rho} into Eq.~\eqref{eq:rho_einstein}, we obtain
\begin{equation}
-\frac{\beta'(r)}{r^2}
=
8\pi\left(
-\frac{\alpha}{r^4}
+\frac{\eta}{r^7}
\right),
\end{equation}
which immediately gives
\begin{equation}
\beta'(r)
=
\frac{8\pi\alpha}{r^2}
-
\frac{8\pi\eta}{r^5}.
\label{eq:beta_prime}
\end{equation}
Integrating Eq.~\eqref{eq:beta_prime} from $r_0$ to $r$ and imposing the throat condition $\beta(r_0)=r_0$, one obtains
\begin{equation}
\beta(r)
=
r_0
+
8\pi\alpha\left(\frac{1}{r_0}-\frac{1}{r}\right)
+
2\pi\eta\left(\frac{1}{r^4}-\frac{1}{r_0^4}\right).
\label{eq:beta}
\end{equation}
This expression determines the shape function, and hence the radial spatial geometry, in terms of the throat radius and the parameters characterizing the Casimir and memory sectors. In the limit $\eta\rightarrow0$, Eq.~\eqref{eq:beta} reduces to the pure Casimir configuration.

Evaluating Eq.~\eqref{eq:beta_prime} at the throat gives
\begin{equation}
\beta'(r_0)
=
\frac{8\pi\alpha}{r_0^2}
-
\frac{8\pi\eta}{r_0^5}.
\label{eq:beta_throat}
\end{equation}
Therefore, the hyperbolic flaring-out condition $\beta'(r_0)>1$ becomes
\begin{equation}
\frac{8\pi\alpha}{r_0^2}
-
\frac{8\pi\eta}{r_0^5}
>
1.
\label{eq:flareout}
\end{equation}
Equivalently, the local flaring-out domain can be written as
\begin{equation}
0\leq\eta<\eta_{\rm crit},
\qquad
\eta_{\rm crit}=\alpha r_0^3-\frac{r_0^5}{8\pi},
\label{eq:eta_crit}
\end{equation}
provided $\eta_{\rm crit}>0$. This relation determines the parameter domain compatible with the local throat flaring-out condition. It should not by itself be interpreted as the complete global admissibility domain, since the Euclidean-embedding condition in Eq.~\eqref{eq:embedding_condition} must be checked separately when an embedding diagram is required. Positive values of $\eta$ reduce $\beta'(r_0)$ and consequently make the flaring-out condition more restrictive. In the pure Casimir limit, $\eta=0$, Eq.~\eqref{eq:flareout} reduces to
\begin{equation}
\frac{8\pi\alpha}{r_0^2}>1,
\end{equation}
which provides a lower bound on the Casimir parameter required to satisfy the local hyperbolic flaring-out condition.
\begin{figure}[ht]
    \centering
    \includegraphics[width=0.6\textwidth]{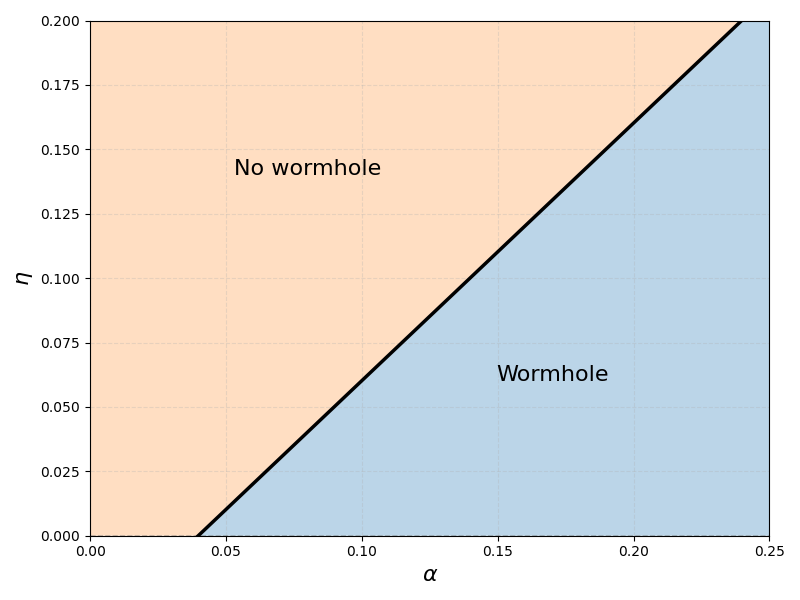}
    \caption{
Local flaring-out domain in the $(\alpha,\eta)$ plane for a fixed throat radius $r_0=1$. The solid line corresponds to the critical value $\eta_{\rm crit}$ obtained from $\beta'(r_0)>1$. Points below the curve satisfy the throat condition $0\leq\eta<\eta_{\rm crit}$, while points above it do not. This figure is intentionally a map of the local throat criterion only; the global Euclidean-embedding bound $\beta(r)\leq2r$, Eq.~\eqref{eq:embedding_condition}, is a separate requirement and is not encoded in this panel.
}
    \label{fig:stability_region}
\end{figure}

The analytical shape function \eqref{eq:beta} provides the radial geometry associated with the prescribed Casimir-memory density profile. In the next section, this result will be combined with Herrera's vanishing-complexity condition in order to determine the redshift function and complete the wormhole solution.

\section{Vanishing Complexity, Redshift Function, and Smooth Matching}
\label{sec:complexity}

In the previous section, the Casimir-memory-like density profile completely determined the shape function through the Einstein field equations \eqref{eq:rho_einstein}. However, the spacetime geometry remains underdetermined since the redshift function $\Phi(r)$ is still unknown. To close the system, we adopt the vanishing-complexity condition, which provides an additional geometrical constraint relating the metric functions. This approach allows the redshift function to be obtained directly from the complexity sector, without the need to impose an explicit equation of state. Once the interior geometry is fully specified, we investigate the possibility of smoothly matching the wormhole spacetime to an exterior hyperbolic Schwarzschild solution through Darmois junction conditions. We then relax exact vanishing complexity in a controlled way through a one-parameter deformation. Besides probing how residual complexity affects the temporal geometry and the matching surface, this deformation will provide the link to the local-complexity prescription developed in Sec.~\ref{sec:minimal_complexity}.

\subsection{Vanishing complexity and redshift function}

A natural framework for closing the system is provided by Herrera's concept of complexity. Originally introduced in the study of self-gravitating anisotropic fluids, the complexity factor measures the combined effects of density inhomogeneity and pressure anisotropy. More recently, Avalos \emph{et al.} \cite{Avalos:2025hfw} extended this concept to hyperbolically symmetric wormhole geometries and showed that the resulting complexity factor provides a useful characterization of Casimir-supported wormholes.

In the present work, rather than evaluating the complexity associated with a given solution, we restrict our attention to the subclass of configurations satisfying the vanishing-complexity condition
\begin{equation}
Y_{TF}=0.
\label{eq: Y_TF}
\end{equation}
From a physical point of view, this condition does not imply either isotropy or homogeneity. Instead, it corresponds to a precise balance between the contributions of density inhomogeneity and pressure anisotropy. Consequently, it provides a physically motivated closure relation that allows the redshift function to emerge dynamically once the matter distribution has been specified.

For hyperbolic wormholes, the complexity factor $Y_{TF}$ obtained by Avalos \emph{et al.} for hyperbolically symmetric configurations is
\begin{equation}
Y_{TF}=\left(\frac{\beta}{r}-1\right)
\left(\Phi''+\Phi'^2\right)
+\frac{\Phi'}{r}
\left(
\frac{\beta'}{2}
-\frac{3\beta}{2r}
+1
\right)
+\frac{1}{2r^3}
\left[
r_0\beta'(r_0)-3\beta(r_0)
\right].
\label{eq:YTF}
\end{equation}

Using the throat condition $\beta(r_0)=r_0$, Eq.~\eqref{eq:YTF} becomes
\begin{equation}
\left(\frac{\beta}{r}-1\right)
\left(\Phi''+\Phi'^2\right)
+\frac{\Phi'}{r}
\left(
\frac{\beta'}{2}
-\frac{3\beta}{2r}
+1
\right)
+\frac{r_0}{2r^3}
\left[
\beta'(r_0)-3
\right]
=0.
\label{eq:complexity_eq}
\end{equation}
Equation~\eqref{eq:complexity_eq} constitutes a nonlinear second-order differential equation for the redshift function. Since both \(\beta(r)\) and \(\beta'(r_0)\) are already known from
Eqs.~\eqref{eq:beta} and \eqref{eq:beta_throat}, respectively,
the vanishing-complexity condition determines the temporal component of the
metric once the source parameters, throat regularity, and temporal
normalization are specified.

For numerical purposes, it is convenient to introduce the auxiliary function
\begin{equation}
u(r)=e^{\Phi(r)}.
\end{equation}
Using
\begin{equation}
\Phi''+\Phi'^2=\frac{u''}{u},
\end{equation}
the complexity equation can be rewritten as
\begin{equation}
\left(\frac{\beta}{r}-1\right)u''
+
\frac{1}{r}
\left(
\frac{\beta'}{2}
-\frac{3\beta}{2r}
+1
\right)u'
+
\frac{r_0}{2r^3}
\left[
\beta'(r_0)-3
\right]u=0.
\label{eq:u_equation}
\end{equation}
This linear second-order equation provides a convenient starting point for the numerical integration of the redshift function.

Regularity at the throat imposes a nontrivial relation between $u(r_0)$ and $u'(r_0)$. Since the overall normalization of the lapse function can be absorbed into a rescaling of the time coordinate, we choose
\begin{equation}
u(r_0)=1.
\end{equation}
Taking the regular limit of Eq.~\eqref{eq:u_equation} at the throat gives
\begin{equation}
 r_0\frac{u'(r_0)}{u(r_0)}=
 \frac{3-\beta'(r_0)}{\beta'(r_0)-1}.
 \label{eq:throat_regularity_p3}
\end{equation}
With the normalization $u(r_0)=1$, this becomes
\begin{equation}
 u'(r_0)=\frac{3-\beta'(r_0)}{r_0[\beta'(r_0)-1]}.
\end{equation}
Once these initial conditions are specified, the redshift function is uniquely determined throughout the retained interior region.

The resulting solution must remain finite and nonvanishing from the throat up to the matching surface (or, if reached first, the geometric boundary), ensuring the absence of horizons in the retained wormhole interior. The corresponding numerical solutions for the complexity-free configuration will be presented in Sec.~\ref{general-complex}, where they will be compared with the complexity deformation introduced below.

\subsection{Matching Conditions}

Having obtained the redshift function, the next step is to construct a complete spacetime by matching the interior wormhole geometry to an exterior vacuum solution. Following Ref.~\cite{Avalos:2025hfw}, we consider a junction surface $\Sigma$ located at $r=r_\Sigma$, where the interior metric
\begin{equation}
ds^2_-=
-e^{2\Phi(r)}dt^2
+\frac{dr^2}{\beta(r)/r-1}
+r^2d\Omega_{(-1)}^2,
\end{equation}
is matched to the hyperbolic Schwarzschild exterior
\begin{equation}
ds^2_+=
-\left(\frac{2M}{r}-1\right)dT^2
+\frac{dr^2}{\frac{2M}{r}-1}
+r^2d\Omega_{(-1)}^2.
\end{equation}

The Darmois matching conditions require the continuity of the first and second fundamental forms across $\Sigma$. For static hyperbolic configurations, these conditions reduce to the continuity of the metric functions together with
\begin{equation}
p_r(r_\Sigma)=0.
\label{eq:darmois_pressure}
\end{equation}

Since the redshift function is obtained numerically, the matching radius must also be determined numerically. We define it as the first regular zero of the radial pressure encountered after the throat and before the first geometric boundary $B(x)=x$, requiring throughout the retained interval that the lapse amplitude satisfies $u>0$ and that the static hyperbolic factor satisfies $B(x)/x-1>0$. A sign change across $u=0$ is therefore treated as a pressure pole rather than as a matching surface.

To investigate the existence of smooth junctions, we introduce the radial
coordinate and source parameters
\begin{equation}
x=\frac{r}{r_0},
\qquad
x_\Sigma=\frac{r_\Sigma}{r_0},
\qquad
a=\frac{8\pi\alpha}{r_0^2},
\qquad
e=\frac{8\pi\eta}{r_0^5}.
\label{eq:dimensionless_parameters}
\end{equation}
We then performed a numerical scan in the \((a,e)\) parameter space over the complete static interval between the throat and the first geometric boundary, without imposing an auxiliary upper cutoff on $x_\Sigma$.
For the representative configuration
\begin{equation}
a=4,
\qquad
e=0.5,
\end{equation}
the vanishing-complexity solution yields
\begin{equation}
r_\Sigma\simeq2.04310r_0.
\label{eq:matching_radius_p3}
\end{equation}
The continuity of the radial metric component implies
\begin{equation}
\frac{\beta(r_\Sigma)}{r_\Sigma}-1=
\frac{2M}{r_\Sigma}-1,
\end{equation}
which leads to
\begin{equation}
M=\frac{\beta(r_\Sigma)}{2}.
\end{equation}

Evaluating the shape function at the matching surface gives
\begin{equation}
M\simeq1.46218r_0.
\end{equation}
Furthermore,
\begin{equation}
\beta(r_\Sigma)-r_\Sigma
\simeq0.88126>0,
\end{equation}
and
\begin{equation}
\frac{2M}{r_\Sigma}-1
\simeq0.43134>0.
\end{equation}

These results show that the matching surface belongs to the admissible hyperbolic region and that the exterior vacuum geometry is well defined at the junction radius.

The continuity of the temporal metric component can always be achieved through a constant rescaling of the exterior time coordinate,
\begin{equation}
T\rightarrow\lambda T,
\end{equation}
which leaves the physical content of the solution unchanged. Since the matching radius is determined by the condition $p_r(r_\Sigma)=0$, the second fundamental form is also continuous, implying the absence of surface stresses and eliminating the need for a thin shell.

It is worth noting that the Darmois conditions are fully satisfied at the matching surface. The continuity of the induced metric requires
\begin{equation}
\beta(r_\Sigma)=2M,
\end{equation}
which immediately guarantees the continuity of the angular components of the extrinsic curvature,
\begin{equation}
K_{\theta\theta}^{-}=K_{\theta\theta}^{+},
\qquad
K_{\phi\phi}^{-}=K_{\phi\phi}^{+}.
\end{equation}
Furthermore, the vanishing of the radial pressure at the junction surface,
\begin{equation}
p_r(r_\Sigma)=0,
\end{equation}
together with the interior Einstein equation
\begin{equation}
8\pi p_r=
\frac{\beta}{r^3}
+\frac{2\Phi'}{r}
\left(\frac{\beta}{r}-1\right),
\end{equation}
implies
\begin{equation}
\Phi'_{\rm int}(r_\Sigma)
=
-\frac{M}
{r_\Sigma^2\left(2M/r_\Sigma-1\right)},
\end{equation}
which coincides exactly with the corresponding exterior expression. Therefore, the temporal component of the extrinsic curvature is also continuous,
\begin{equation}
K_{tt}^{-}=K_{tt}^{+},
\end{equation}
and the junction is a genuine Darmois matching, requiring no thin shell or surface stresses.

We conclude that the vanishing-complexity hyperbolic wormhole solutions admit smooth Darmois matchings to a hyperbolic vacuum exterior over an extended region of the parameter space. Nevertheless, the existence of a wormhole throat alone does not guarantee a shell-free matching. In particular, the pure Casimir limit $\eta=0$ may satisfy the flare-out condition and still fail to develop a regular zero of the radial pressure before the first geometric boundary, in which case a shell-free Darmois junction is not obtained within the present construction.

\subsection{Complexity deformation}
\label{general-complex}

Having established the existence of smooth Darmois matchings in the vanishing-complexity sector, we now investigate the robustness of this result under a one-parameter deformation of the complexity equation. 

The complexity-free configuration considered throughout this work is governed by
\begin{equation}\label{eq:p3}
\left(\frac{\beta}{r}-1\right)u''+
\frac{1}{r}
\left(
\frac{\beta'}{2}
-\frac{3\beta}{2r}
+1
\right)u'
+
\frac{r_0}{2r^3}
\left[
\beta'(r_0)-3
\right]
u
=0,
\end{equation}
where $u(r)=e^{\Phi(r)}$, as defined previously.

We now promote the constant coefficient that appears in the term associated with the throat to a free deformation parameter $p$ through
\begin{equation}
\beta'(r_0)-3
\longrightarrow
\beta'(r_0)-p.
\end{equation}
The resulting equation becomes
\begin{equation}\label{eq:pgeneral}
\left(\frac{\beta}{r}-1\right)u''+
\frac{1}{r}
\left(
\frac{\beta'}{2}
-\frac{3\beta}{2r}
+1
\right)u'
+
\frac{r_0}{2r^3}
\left[
\beta'(r_0)-p
\right]
u
=0,
\end{equation}
with the original vanishing-complexity configuration being recovered for $p=3$. Substituting Eq.~\eqref{eq:pgeneral} into the original expression \eqref{eq:YTF} shows that the deformation has a direct complexity interpretation:
\begin{equation}
Y_{TF}(r)=\frac{(p-3)r_0}{2r^3}.
\label{eq:YTF_pdeformation}
\end{equation}
Thus, $p=3$ gives $Y_{TF}=0$, whereas $p\neq3$ prescribes a residual complexity profile decaying as $r^{-3}$, with $p-3$ controlling its amplitude. The deformation parameter therefore quantifies departures from the vanishing-$Y_{TF}$ sector and provides a convenient way to investigate the impact of residual complexity on the wormhole geometry and matching properties.

For the branch admitting a finite derivative $u'(r_0)$ in the areal radial coordinate, the throat limit of Eq.~\eqref{eq:pgeneral} gives
\begin{equation}
r_0\frac{u'(r_0)}{u(r_0)}=
\frac{p-\beta'(r_0)}{\beta'(r_0)-1}.
\label{eq:throat_regularity_p}
\end{equation}
This condition characterizes the finite-$r$-derivative branch of the local solution at the throat. As discussed in Sec.~V, Eq.~\eqref{eq:pgeneral} also admits a second local branch whose derivative with respect to the areal coordinate $r$ is singular at $r=r_0$, but which is regular when expressed in terms of the proper radial distance.

While the density profile and shape function remain unchanged, the temporal component of the metric and the matching properties of the resulting spacetime become sensitive to the value of $p$, generating novel families of wormhole solutions.
\begin{figure}[!htbp]
\begin{minipage}{0.48\textwidth}
    \centering
    \includegraphics[width=\textwidth]{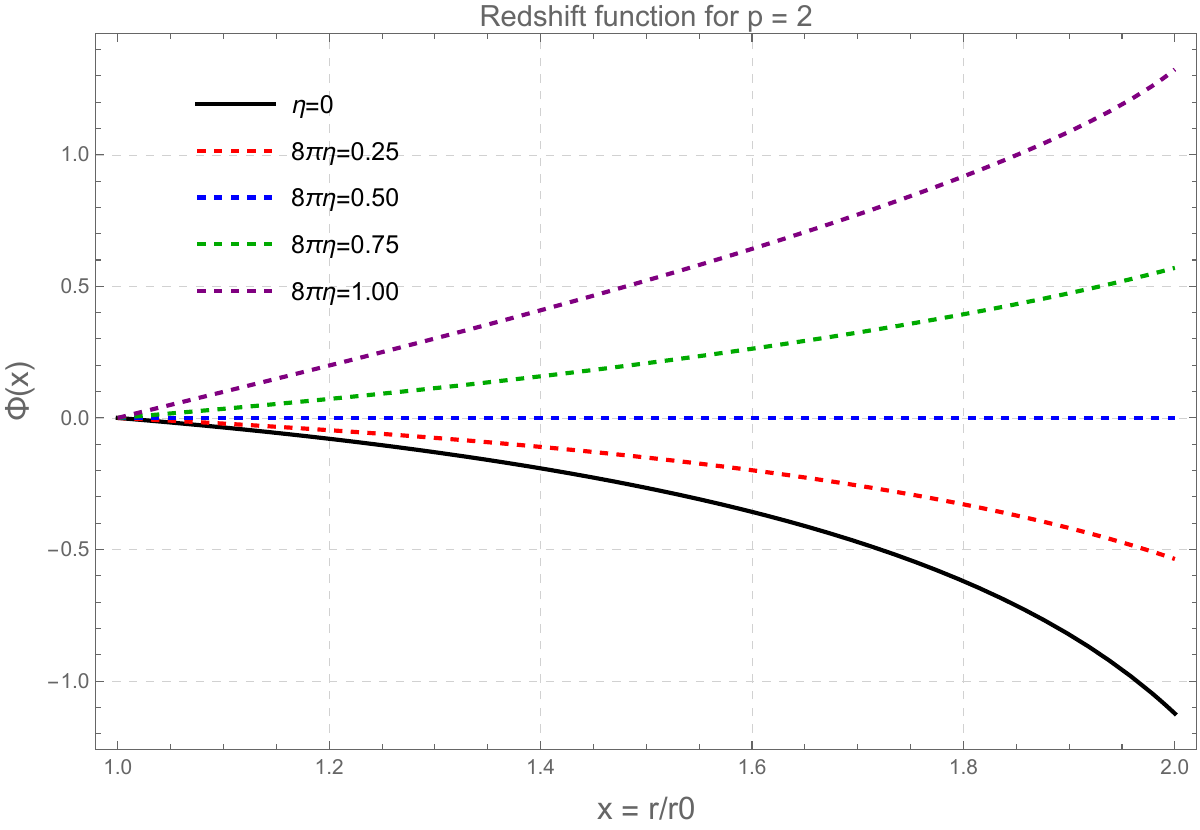}
\end{minipage}
\hfill
\begin{minipage}{0.48\textwidth}
    \centering
    \includegraphics[width=\textwidth]{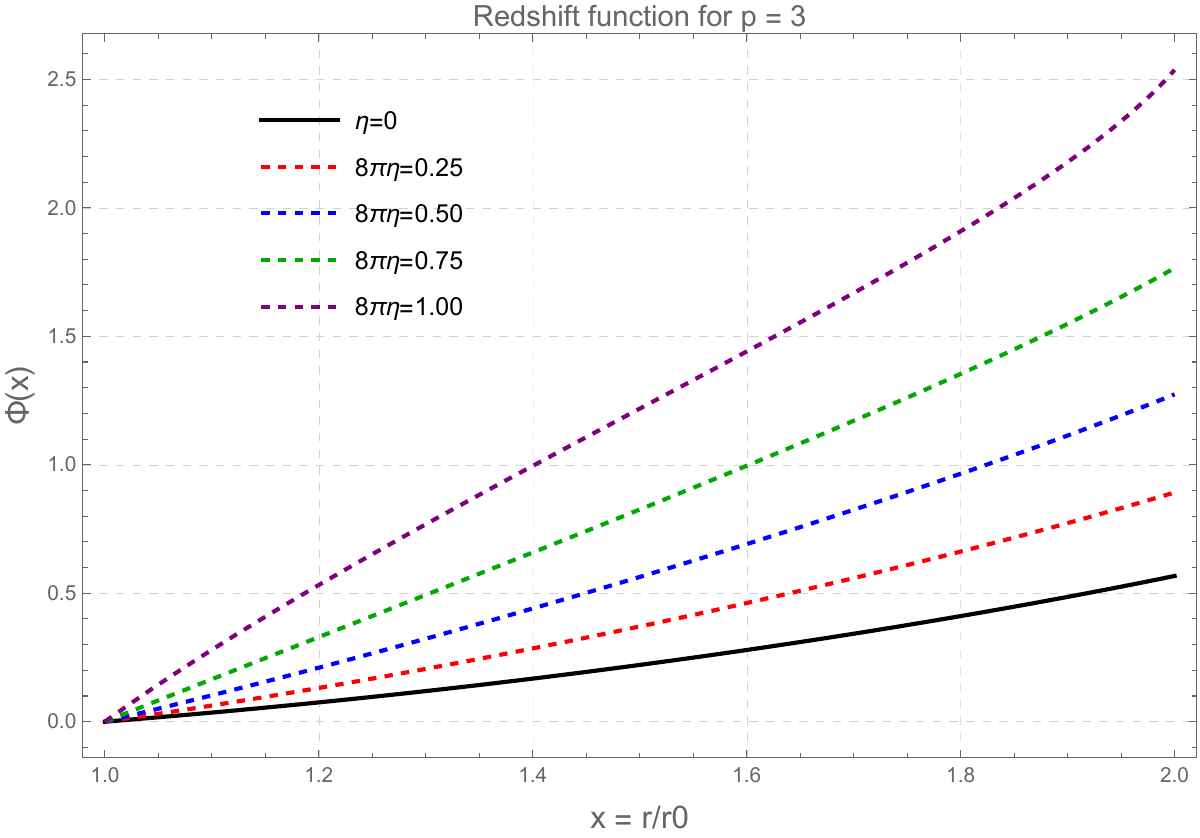}
\end{minipage}
\caption{
Redshift function $\Phi(x)$, with $x=r/r_0$, obtained from the $p$-deformed complexity equation, with $r_0=1$ and $8\pi\alpha/r_0^2=2.5$, for $p=2$ (upper panel) and $p=3$ (lower panel). The solid curve corresponds to the pure Casimir case ($\eta=0$), while the dashed curves represent memory-corrected configurations with increasing values of $8\pi\eta/r_0^5$. For the representative values of $p$ and the source parameters shown, increasing the memory parameter enhances the redshift profile.
}
\label{fig:redshift_memory}
\end{figure}
In fact, Figure~\ref{fig:redshift_memory} shows the numerical solutions obtained for two representative values of the deformation parameter. In both panels, the pure Casimir configuration corresponds to the lowest curve, while increasing values of the memory parameter produce progressively larger values of the redshift function. Thus, for the representative values of $p$ and source parameters shown, the gravitational-memory contribution enhances the temporal component of the metric. Although some $p=2$ solutions develop negative values of $\Phi(r)$, the corresponding lapse function remains strictly positive, and therefore no horizons are formed.

A particularly interesting situation occurs when
\begin{equation}
p=\beta'(r_0).
\label{eq:p_endpoint}
\end{equation}
For the finite-$r$-derivative branch, Eq.~\eqref{eq:throat_regularity_p} then gives $u'(r_0)=0$, while the last term of Eq.~\eqref{eq:pgeneral} vanishes identically. The constant solution
\begin{equation}
u(r)=\mathrm{const.},
\qquad
\Phi'(r)=0,
\end{equation}
is therefore admitted globally. This corresponds to a constant-redshift (zero-tidal-force) wormhole and may be interpreted as a balance between the complexity deformation and the throat geometry. Importantly, however, at the distinguished value \eqref{eq:p_endpoint} the constant-redshift configuration is not the most general solution of Eq.~\eqref{eq:pgeneral}. A second independent branch is also admitted. Its relation to the locally vanishing-complexity condition and its regularity at the throat are analyzed in Sec.~V.

The complexity deformation also affects the location of the matching surface. For the same representative configuration,
\begin{equation}
a=4,
\qquad
e=0.5,
\end{equation}
the first regular root of the radial pressure occurs at
\begin{equation}
r_\Sigma\simeq1.38372r_0,
\qquad
(p=2),
\label{eq:matching_radius_p2}
\end{equation}
whereas the original complexity-free configuration yields
\begin{equation}
r_\Sigma\simeq2.04310r_0,
\qquad
(p=3).
\end{equation}
For this representative configuration, decreasing the deformation parameter from $p=3$ to $p=2$ moves the first regular matching surface closer to the throat.

For the case $p=2$, the continuity condition gives
\begin{equation}
M=\frac{\beta(r_\Sigma)}{2}
\simeq1.00917r_0.
\end{equation}
Moreover,
\begin{equation}
\frac{\beta(r_\Sigma)}{r_\Sigma}-1=
\frac{2M}{r_\Sigma}-1
\simeq0.45863>0,
\end{equation}
while
\begin{equation}
u(r_\Sigma)\simeq0.72684.
\end{equation}
The solution remains regular throughout the interval
\begin{equation}
u(r)>0,
\qquad
r_0<r<r_\Sigma,
\end{equation}
demonstrating that the deformed configuration also admits a smooth Darmois matching to the hyperbolic Schwarzschild exterior.
\begin{figure}[!htbp]
\begin{minipage}{0.48\textwidth}
    \centering
    \includegraphics[width=\linewidth]{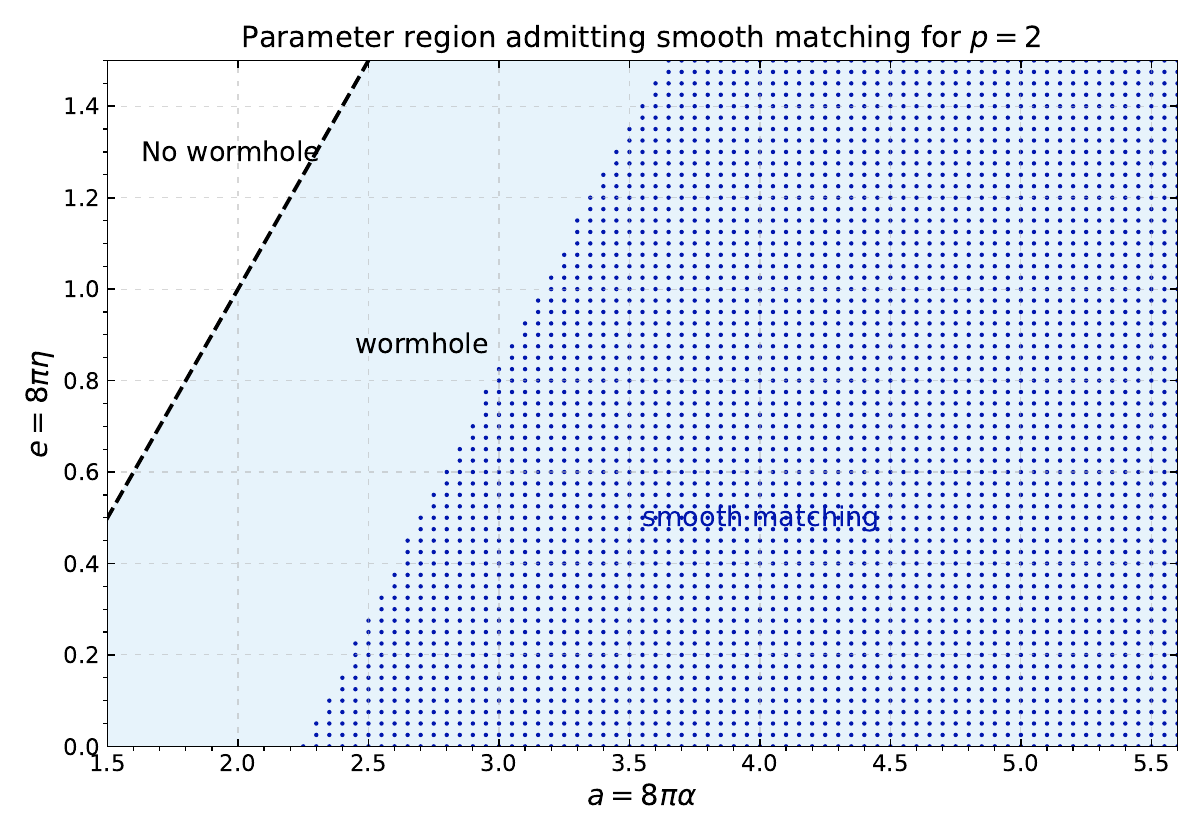}
\end{minipage}
\hfill
\begin{minipage}{0.48\textwidth}
    \centering
    \includegraphics[width=\linewidth]{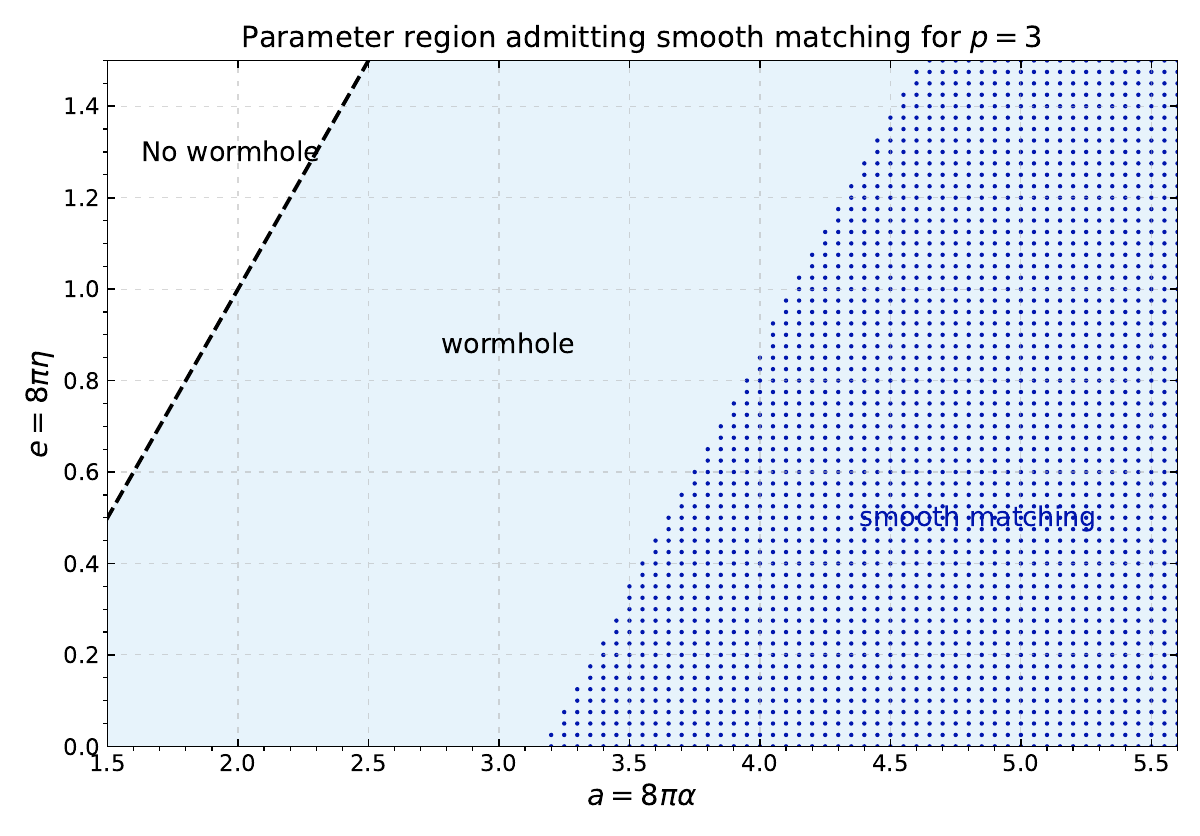}
\end{minipage}
\caption{Parameter region admitting smooth matching to a hyperbolic vacuum exterior for $p=2$ (upper panel) and $p=3$ (lower panel), with $r_0=1$. The pale-blue sector satisfies the hyperbolic flare-out condition, and the black dashed line is its threshold $a-e=1$. The blue points identify the sampled parameters for which the first regular zero of $p_r$ occurs while $u>0$ and $B(x)/x-1>0$. The search extends from the throat to the first geometric boundary $B(x)=x$, with no auxiliary upper cutoff on $x_\Sigma$. Over the displayed range $a\leq5.5$ and $e\geq0$, the Euclidean-embedding upper bound $B(x)\leq2x$ is automatically satisfied for all $x\geq1$.}
\label{fig:matching_region}
\end{figure}
The full-domain scan in Fig.~\ref{fig:matching_region} confirms that the $p=2$ and $p=3$ families admit shell-free Darmois junctions over finite portions of the displayed parameter space. At every sampled value of $e$, however, the threshold value of $a$ is lower for $p=2$ than for $p=3$. Hence the deformed $p=2$ family possesses the larger matching domain within the range shown. This comparison is consistent with the representative radii above, for which decreasing $p$ moves the first pressure-free surface closer to the throat.

The blue points already incorporate the regularity tests rather than representing every numerical zero of $p_r$. Candidate roots were accepted only before the first loss of either the positive lapse or the static hyperbolic factor, and a lapse zero was rejected as a pressure pole. Although the upper Euclidean-embedding condition was not required as an independent numerical cutoff in this scan, it is automatically fulfilled throughout the displayed source domain. Indeed, for $e\geq0$ the memory term lowers $B(x)$ relative to the pure-Casimir case, and for $e=0$ one has $B(x)\leq2x$ for all $x\geq1$ whenever $a\leq3+2\sqrt{2}\simeq5.828$, which contains the plotted range $a\leq5.5$. Consequently, the difference between the two panels measures the dependence of the regular shell-free matching domain on the complexity deformation, not merely the occurrence of an unconstrained sign change in the radial pressure.

For the representative configuration $a=4$ and $e=0.5$, a bisection search over the complete static interval gives $p_{\max}\simeq3.3718$. Regular roots remain present at $p=3.20$ and $p=3.30$ and approach the limiting configuration as $p\to p_{\max}^{-}$. The earlier apparent loss of roots near $p\simeq3.20$ was therefore caused by a finite radial search window and does not represent the physical upper limit.

\begin{figure}[!htbp]
\centering
\includegraphics[width=0.7\textwidth]{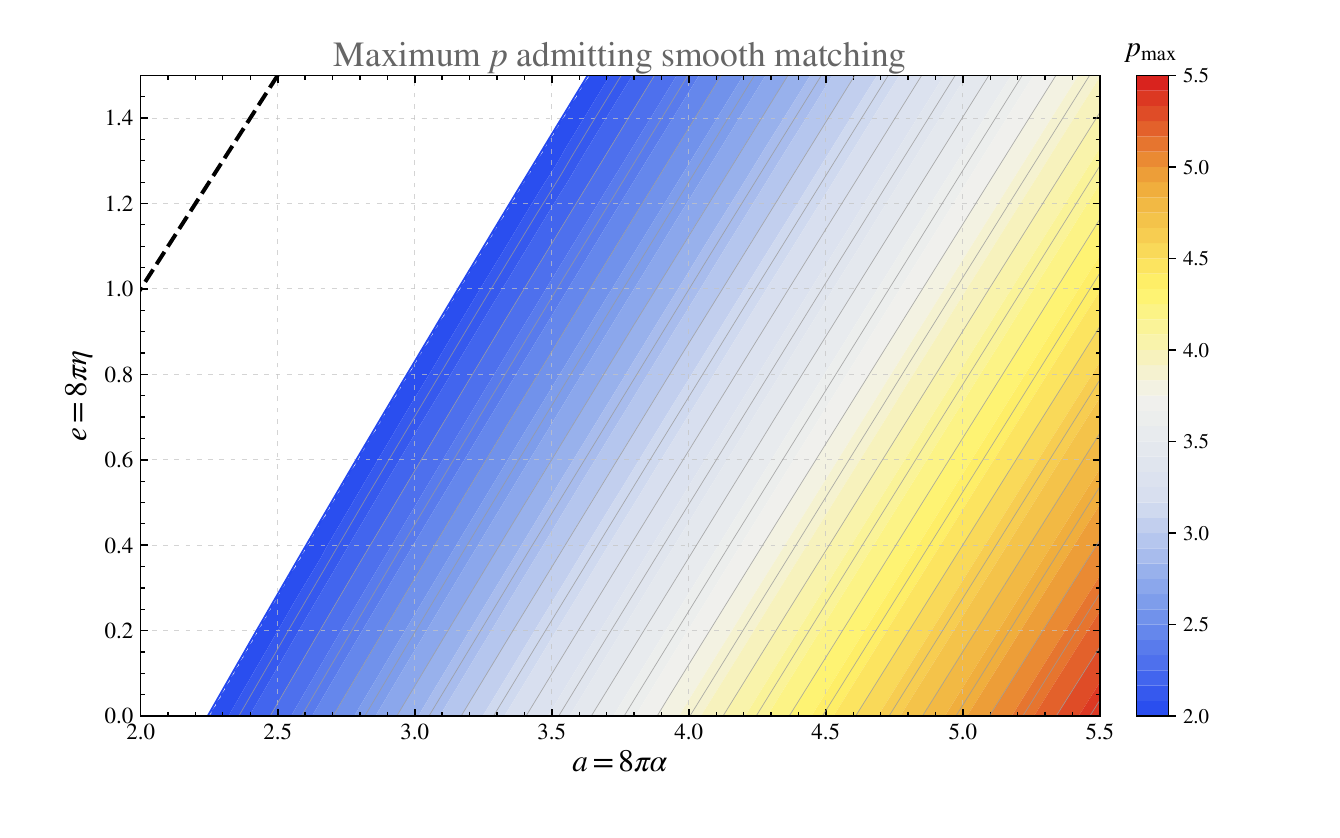}
\caption{
Maximum value of the complexity-deformation parameter $p$ admitting a regular first pressure-free Darmois surface over the full static interval, for $r_0=1$ and the search range $2\leq p\leq6$. The color scale gives $p_{\max}$ at each sampled point of the $(a,e)$ parameter space. Blank points do not admit a regular matching root already at $p=2$, and the black dashed line is the flare-out threshold $a-e=1$. Roots are accepted only while $u>0$ and $B(x)/x-1>0$, before the first geometric boundary $B(x)=x$; no auxiliary radial cutoff is imposed. The upper Euclidean-embedding condition $B(x)\leq2x$ is automatically satisfied throughout the displayed range $a\leq5.5$, $e\geq0$.
}
\label{fig:pmax}
\end{figure}

To determine this dependence throughout the displayed source domain, we searched $2\leq p\leq6$ at each point of the $(a,e)$ grid and identified the supremum admitting a regular first pressure root before $B(x)=x$. The resulting full-domain distribution is shown in Fig.~\ref{fig:pmax}. Blank points fail the matching test already at $p=2$. Within the scanned range, $p_{\max}$ increases with $a$ at fixed $e$ and decreases with $e$ at fixed $a$; thus the Casimir and memory contributions shift the admissible deformation range in opposite directions.

The nearly parallel contours in Fig.~\ref{fig:pmax} show that the two source parameters primarily translate the limiting deformation across the displayed plane rather than producing disconnected admissible islands. At fixed memory parameter, increasing the Casimir contribution enlarges the interval of $p$ compatible with a regular pressure-free surface. At fixed $a$, increasing $e$ reduces that interval, consistently with the decrease of the throat derivative $a-e$. These trends refer to the finite parameter and deformation ranges shown and do not establish an unrestricted bound outside the scan.

The numerical upper limit on $p$ also admits a useful analytical interpretation. Let $q\equiv\beta'(r_0)>1$. For the finite-$r$-derivative branch, Eq.~\eqref{eq:throat_regularity_p} gives $u'(r_0)>0$ whenever $p>q$. If $u'$ were to vanish for the first time at some point in the static region while $u>0$, Eq.~\eqref{eq:pgeneral} would give
\begin{equation}
\left(\frac{\beta}{r}-1\right)u''
=
\frac{r_0(p-q)}{2r^3}\,u
>0,
\end{equation}
which precludes a crossing of $u'$ from positive to negative values. Hence $u'(r)>0$ throughout the regular static interval for $p>q$.

The limiting case $p=q$ must be treated separately. In this case the last term of Eq.~\eqref{eq:pgeneral} vanishes, and the equation becomes a homogeneous first-order equation for $u'$. The finite-$r$-derivative condition $u'(r_0)=0$ therefore selects the constant solution $u'\equiv0$. Consequently, for all $p\geq q$ in this branch,
\begin{equation}
8\pi p_r=
\frac{\beta}{r^3}
+
\frac{2}{r}
\left(\frac{\beta}{r}-1\right)\frac{u'}{u}
>0
\end{equation}
throughout the static region. A pressure-free Darmois surface thus requires
\begin{equation}
p<\beta'(r_0)=a-e.
\label{eq:p_matching_bound}
\end{equation}
Accordingly, the numerical quantity $p_{\max}$ displayed in Fig.~\ref{fig:pmax} satisfies $p_{\max}\leq\beta'(r_0)$, providing an analytical explanation for the finite deformation range observed in the scan.

There is a further structural feature of the $p$-deformed family that will be important below. Separating the differential part of the complexity factor from the contribution fixed by the nonzero lower boundary at the throat, Eq.~\eqref{eq:pgeneral} implies
\begin{equation}
Y_{TF}^{\mathrm{loc}}(r)
=
\frac{\left[p-\beta'(r_0)\right]r_0}{2r^3}.
\label{eq:YTF_local_p}
\end{equation}
Consequently,
\begin{equation}
Y_{TF}^{\mathrm{loc}}=0
\qquad\Longleftrightarrow\qquad
p=\beta'(r_0).
\label{eq:minimal_endpoint}
\end{equation}
Thus the locally vanishing-complexity prescription is not disconnected from the $p$-deformed construction. Rather, it occurs at a distinguished value of the deformation parameter. At this value of $p$, the finite-$r$-derivative branch contains the constant-redshift solution identified above, while the second independent branch leads to a nontrivial lapse. The latter provides the semianalytical sector developed in Sec.~V.

\section{Locally Vanishing Complexity: The Second Lapse Branch}
\label{sec:minimal_complexity}

The analysis of Sec.~IV shows that the locally vanishing-complexity condition is not an independent closure disconnected from the $p$-deformed family. Indeed, Eq.~\eqref{eq:YTF_local_p} gives
\begin{equation}
Y_{TF}^{\rm loc}=0
\qquad\Longleftrightarrow\qquad
p=\beta'(r_0).
\label{eq:minimal_p_relation}
\end{equation}
The purpose of the present section is therefore to examine the two local solution branches at this distinguished value of the deformation parameter. As already noted in Sec.~IV, the finite-$r$-derivative branch then contains the constant-redshift solution. Here we retain the second independent lapse branch, which is nontrivial in the areal coordinate and admits a regular description in terms of proper radial distance.

For completeness, the full complexity factor may be decomposed as
\begin{equation}
Y_{TF}=
\underbrace{\left(\frac{\beta}{r}-1\right)\left(\Phi''+\Phi'^2\right)
+\frac{\Phi'}{r}\left(\frac{\beta'}{2}-\frac{3\beta}{2r}+1\right)}_{Y_{TF}^{\rm loc}(r)}
+\underbrace{\frac{1}{2r^3}\left[r_0\beta'(r_0)-3\beta(r_0)\right]}_{Y_{TF}^{\rm th}(r)}.
\label{eq:YTF_split}
\end{equation}
Using $\beta(r_0)=r_0$, the second contribution is fixed entirely by throat data. Imposing only
\begin{equation}
Y_{TF}^{\rm loc}=
\left(\frac{\beta}{r}-1\right)\left(\Phi''+\Phi'^2\right)
+\frac{\Phi'}{r}\left(\frac{\beta'}{2}-\frac{3\beta}{2r}+1\right)=0
\label{eq:YTF_local_zero}
\end{equation}
therefore retains the throat contribution while eliminating the local differential part. In view of Eq.~\eqref{eq:minimal_p_relation}, this prescription corresponds precisely to the $p=\beta'(r_0)$ sector of Eq.~\eqref{eq:pgeneral}.

Introducing
\begin{equation}
u(r)=e^{\Phi(r)},\qquad \Delta(r)=\beta(r)-r,
\end{equation}
Eq.~\eqref{eq:YTF_local_zero} can be integrated by first writing
\begin{equation}
u''+\left(\frac{1}{2}\frac{\Delta'}{\Delta}-\frac{3}{2r}\right)u'=0.
\label{eq:minimal_u_linear}
\end{equation}
For the nonconstant branch, division by $u'$ gives
\begin{equation}
\frac{u''}{u'}=-\frac{1}{2}\frac{\Delta'}{\Delta}+\frac{3}{2r},
\label{eq:minimal_u_ode}
\end{equation}
and hence
\begin{equation}
u(r)=C_1+C_2\int_{r_0}^{r}
\frac{\bar r^{3/2}}{\sqrt{\beta(\bar r)-\bar r}}\,d\bar r.
\label{eq:u_general_solution}
\end{equation}
The constant solution is recovered for $C_2=0$ and coincides with the finite-$r$-derivative branch identified in Sec.~IV. The genuinely nonconstant solution has $C_2\neq0$. Choosing the interior time normalization $u(r_0)=1$ gives $C_1=1$, so that
\begin{equation}
u(r)=e^{\Phi(r)}=1+C_2 I(r),
\end{equation}
where
\begin{equation}
I(r)=\int_{r_0}^{r}
\frac{\bar r^{3/2}}{\sqrt{\beta(\bar r)-\bar r}}\,d\bar r.
\label{eq:I_def}
\end{equation}

At first sight the nonconstant branch appears singular because
$I'(r)\sim(r-r_0)^{-1/2}$ at a regular flaring throat. This behavior is, however, a consequence of the areal radial coordinate. Introduce the signed proper radial distance $\ell$ through
\begin{equation}
d\ell=\frac{dr}{\sqrt{\beta/r-1}},
\qquad
\left(\frac{dr}{d\ell}\right)^2=\frac{\beta}{r}-1.
\label{eq:proper_distance_minimal}
\end{equation}
Since
\begin{equation}
I'(r)=\frac{r^{3/2}}{\sqrt{\beta-r}},
\end{equation}
one obtains on either side of the throat
\begin{equation}
\frac{dI}{d\ell}=\pm r,
\qquad
\frac{du}{d\ell}=\pm C_2 r.
\label{eq:minimal_proper_derivative}
\end{equation}
Thus $du/d\ell$ remains finite at $r=r_0$, despite the divergence of $du/dr$. Moreover, differentiating Eq.~\eqref{eq:proper_distance_minimal} gives
\begin{equation}
\left.\frac{d^2r}{d\ell^2}\right|_{r_0}
=
\frac{\beta'(r_0)-1}{2r_0},
\end{equation}
while Eq.~\eqref{eq:minimal_proper_derivative} implies
\begin{equation}
\left.\frac{d^2u}{d\ell^2}\right|_{r_0}=0.
\end{equation}
This conclusion can also be checked directly at the level of curvature. For the smooth proper-distance extension, the Kretschmann scalar remains finite at the throat,
\begin{equation}
K(r_0)=
\frac{2[\beta'(r_0)-1]^2+4}{r_0^4}.
\label{eq:K_throat_minimal}
\end{equation}
Thus the divergence of $du/dr$ is purely associated with the areal radial coordinate and does not signal a curvature singularity.

For a smooth two-sided extension of the nonconstant branch, the outward-coordinate integration constants on the two sides must be chosen with opposite signs, $C_{2,-}=-C_{2,+}$. With this choice, $u(\ell)$ is differentiable through $\ell=0$ and locally behaves as
\begin{equation}
u(\ell)=u_0+C\,r_0\ell+O(\ell^3).
\end{equation}
Using the same sign of $C_2$ on two reflected copies would instead generate a term proportional to $|\ell|$ and hence a distributional layer at the throat. The smooth nonconstant extension is therefore generally not reflection symmetric in the lapse; a fully reflection-symmetric smooth extension selects $C_2=0$. In what follows we work on one matter-supported side, $r_0\leq r\leq R$, for which the matching construction is unchanged.

We retain the same Casimir-memory density and shape function introduced in Sec.~\ref{sec:shape}, Eqs.~\eqref{eq:rho} and~\eqref{eq:beta}, and therefore the same hyperbolic flaring-out requirement $\beta'(r_0)>1$, or equivalently $a-e>1$ in terms of the dimensionless parameters of Eq.~\eqref{eq:dimensionless_parameters}. No new matter ansatz is introduced in this sector; the nonconstant lapse is the second solution branch of the locally vanishing-complexity equation.

We now impose the Darmois conditions at $r=R\equiv r_\Sigma$. Let
\begin{equation}
f(R)=\frac{2M}{R}-1,
\end{equation}
and allow for the same constant rescaling of the exterior time coordinate used in Sec.~IV, $T=\lambda t$. Continuity of the first fundamental form gives
\begin{equation}
\beta(R)=2M,
\label{eq:minimal_match_beta}
\end{equation}
and
\begin{equation}
u(R)=\lambda\sqrt{f(R)}.
\label{eq:minimal_match_time}
\end{equation}
Continuity of the second fundamental form is equivalent here to the absence of a surface layer, or $p_r(R)=0$, which yields
\begin{equation}
\frac{u'(R)}{u(R)}=-\frac{M}{R^2 f(R)}.
\label{eq:minimal_match_derivative}
\end{equation}
Using $u=C_1+C_2I$ and $I'(R)=R/\sqrt{f(R)}$, Eq.~\eqref{eq:minimal_match_derivative} fixes the ratio of integration constants as
\begin{equation}
\frac{C_2}{C_1}=-\frac{M}{R^3\sqrt{f(R)}+M I(R)}.
\label{eq:minimal_match_ratio}
\end{equation}
After the interior normalization $C_1=1$ is chosen, the general matching relations become
\begin{align}
C_2&=-\frac{M}{R^3\sqrt{f(R)}+M I(R)},\label{eq:minimal_match_C2_general}\\
\lambda&=\frac{R^3}{R^3\sqrt{f(R)}+M I(R)}.\label{eq:minimal_match_lambda}
\end{align}
Thus $C_1=1$ and $\lambda=1$ are distinct normalization choices: the former fixes the interior time scale at the throat, whereas the latter additionally identifies the exterior and interior time normalizations at the junction. The Darmois conditions themselves do not require $\lambda=1$.

A convenient normalized subclass is obtained by imposing the additional choice $\lambda=1$. In this case Eqs.~\eqref{eq:minimal_match_C2_general} and~\eqref{eq:minimal_match_lambda} reduce to
\begin{align}
I(R)&=\frac{R^3}{M}\left(1-\sqrt{f(R)}\right),\label{eq:minimal_match_I_subclass}\\
C_2&=-\frac{M}{R^3},\label{eq:minimal_match_C2_subclass}
\end{align}
and the lapse takes the form
\begin{equation}
e^{\Phi(r)}=1-\frac{M}{R^3}I(r).
\label{eq:minimal_lapse_subclass}
\end{equation}
Because $I(r)$ is monotonically increasing for $r_0<r\leq R$ and $C_2<0$, this normalized subclass has a monotonically decreasing lapse with
\begin{equation}
u(R)=\sqrt{f(R)}>0,
\end{equation}
so no horizon is encountered in the matter-supported region whenever the matching surface lies in the static hyperbolic sector.

The general nonconstant locally vanishing-complexity construction is therefore less restrictive than the normalized subclass: once an admissible matching radius $R$ is selected, Eq.~\eqref{eq:minimal_match_beta} fixes $M$, Eq.~\eqref{eq:minimal_match_C2_general} fixes the remaining interior integration constant, and Eq.~\eqref{eq:minimal_match_lambda} fixes the relative exterior time normalization. The numerical configurations displayed below were obtained with the additional normalization $\lambda=1$ and should be understood as representative members of this particular subclass rather than as an exhaustive characterization of the general nonconstant branch.

For some source parameters, the normalized matching condition above admits more than one admissible root. In the representative configurations displayed below, we select the outermost matching radius. The smaller admissible roots correspond to additional members of the same normalized subclass and do not affect the existence analysis.

\begin{figure}[t!]
    \centering
    \includegraphics[width=0.45\textwidth]{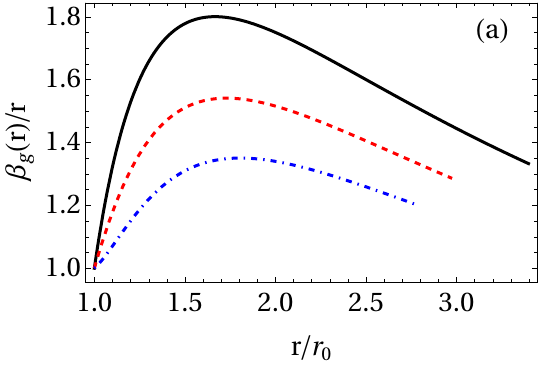}
    \includegraphics[width=0.45\textwidth]{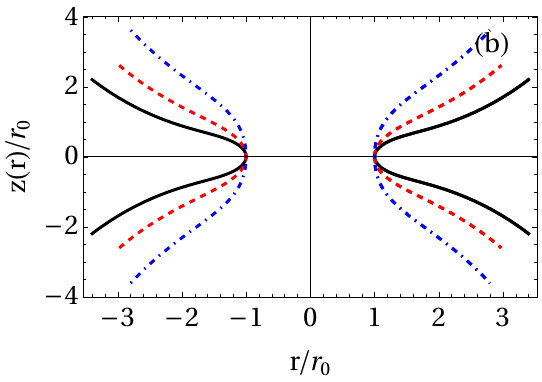}
    \includegraphics[width=0.5\textwidth]{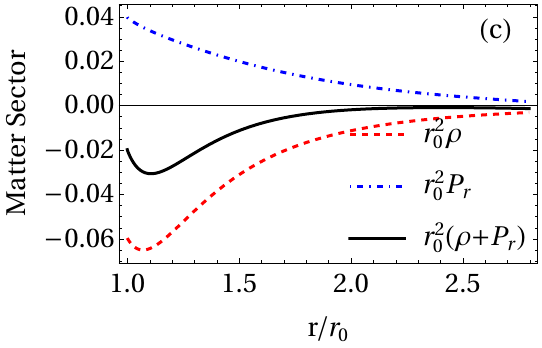}
    \caption{Representative solutions of the normalized nonconstant locally vanishing-complexity branch with $\lambda=1$. (a) $\beta/r$ versus $r/r_0$ for $8\pi\alpha=5r_0^2$ and $8\pi\eta=0$ (black solid, $M=2.26425r_0$, $r_\Sigma=3.39787r_0$), $8\pi\eta=2r_0^5$ (red dashed, $M=1.91309r_0$, $r_\Sigma=2.97585r_0$), and $8\pi\eta=3.5r_0^5$ (blue dot-dashed, $M=1.67564r_0$, $r_\Sigma=2.79639r_0$). (b) Embedding diagram $z/r_0$ for the same configurations. (c) Normalized density $r_0^2\rho$ (red dashed), radial pressure $r_0^2p_r$ (blue dot-dashed), and radial null combination $r_0^2(\rho+p_r)$ (black solid) for $8\pi\alpha=5r_0^2$ and $8\pi\eta=3.5r_0^5$.}
    \label{fig:solution}
\end{figure}

\begin{figure}[t!]
    \centering
    \includegraphics[width=0.6\textwidth]{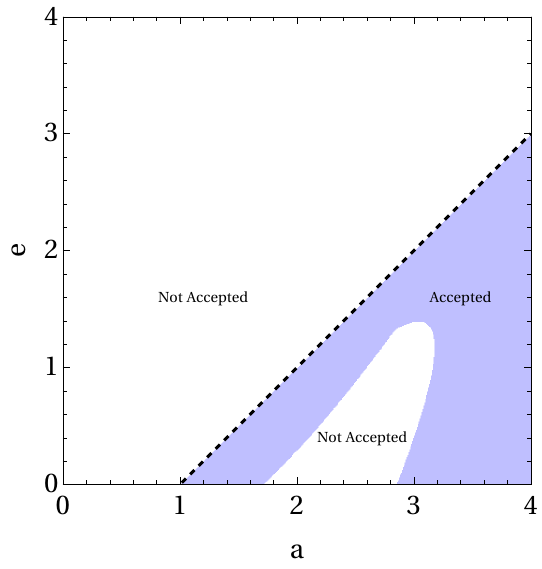}
    \caption{Parameter space of the normalized nonconstant locally vanishing-complexity branch with $\lambda=1$ in the $(a,e)$ plane. The dashed black line marks the flaring-out threshold $a-e=1$; points with $a-e\leq1$ are excluded. Within the admissible throat domain, blue points denote parameter pairs for which at least one matching radius $R=r_\Sigma/r_0$ satisfies the normalized matching condition~\eqref{eq:minimal_match_I_subclass} together with the scan criterion $x<B(x)\leq2x$ for all $x\in(1,R)$, with $B(x)$ defined in Eq.~\eqref{eq:B_dimensionless}. White points inside the flaring-out domain fail at least one of these additional requirements.}
    \label{fig:restrictions}
\end{figure}

To describe the numerical scan in Fig.~\ref{fig:restrictions}, it is useful to define the dimensionless shape function
\begin{equation}
B(x)\equiv\frac{\beta(r_0x)}{r_0},\qquad x=\frac{r}{r_0}.
\label{eq:B_dimensionless}
\end{equation}
According to the geometric discussion in Sec.~\ref{sec:geometry}, the lower bound $B(x)>x$ keeps the solution in the static hyperbolic sector, while the upper bound $B(x)\leq2x$ is the Euclidean-embedding requirement of Eq.~\eqref{eq:embedding_condition}. The numerical scan shown in Fig.~\ref{fig:restrictions} refers specifically to the normalized nonconstant branch with $\lambda=1$, so that the matching radius also satisfies Eq.~\eqref{eq:minimal_match_I_subclass}. Both geometric bounds are imposed throughout the retained interval. A point $(a,e)$ is marked as accepted when at least one such normalized matching radius $R=r_\Sigma/r_0$ exists while
\begin{equation}
x<B(x)\leq2x,\qquad 1<x<R.
\label{eq:minimal_scan_condition}
\end{equation}
The dashed line $a-e=1$ marks the flaring-out threshold; points with $a-e\leq1$ are excluded independently of the matching test.

Figure~\ref{fig:restrictions} shows that, within the additional normalization $\lambda=1$, the flaring-out condition is necessary but not sufficient for the particular subclass represented by the scan. The accepted subset is determined jointly by the source parameters, the normalized matching condition, and the requirement that the spatial section remain Euclidean-embeddable up to the junction. This numerical restriction should not be interpreted as the full existence domain of the general nonconstant locally vanishing-complexity branch, for which the relative time normalization $\lambda$ is instead fixed by Eq.~\eqref{eq:minimal_match_lambda}.

The locally vanishing-complexity sector is therefore an intrinsic distinguished
sector of the $p$-deformed construction rather than an additional independent
closure. It occurs at the distinguished value $p=\beta'(r_0)$, where the finite-$r$-derivative solution is the constant-redshift branch and the second independent solution yields the semianalytical nonconstant lapse developed above. Both lapse branches retain the same Casimir-memory density and shape function. Consequently, the source-dependent energy-condition statements established below apply unchanged, whereas the detailed pressure-dependent analysis of Sec.~VI will continue to refer to the finite-$r$-derivative $p$-family of Sec.~IV.

\section{Energy conditions and their complexity dependence}
\label{sec:energy}

The solution families constructed in the preceding sections share the same
Casimir--memory density profile and the same shape function, but differ in
the temporal geometry selected by their respective complexity prescriptions.
This distinction is particularly useful for the analysis of the energy
conditions. Some conclusions follow solely from the common matter density
and throat geometry and therefore apply to all the complexity branches
considered above. Conditions involving the principal pressures, on the other
hand, depend on the lapse sector. We therefore first isolate the consequences
that are common to the exact vanishing-complexity, \(p\)-deformed, and
nonconstant locally vanishing-complexity lapse sectors, and then restrict the
detailed pressure-dependent analysis to the finite-\(r\)-derivative
\(p\)-deformed family of Eq.~\eqref{eq:pgeneral}, with \(p=3\) recovering
exact vanishing complexity.

The anisotropic energy--momentum tensor \eqref{eq:stress_tensor} is of type I
and, in its orthonormal principal frame, has energy density \(\rho\) and
principal pressures \(p_r\) and \(p_\perp\). This eigenvalue structure allows
the energy conditions to be constructed directly from the energy measured by
causal observers \cite{Visser:1995cc,Lobo:2017cay}.

The null energy condition (NEC) requires
\(T_{\mu\nu}k^\mu k^\nu\geq0\) for every null vector \(k^\mu\). Through
Einstein's equations and the Raychaudhuri equation, this contraction is the
matter contribution to null-geodesic focusing. The weak energy condition
(WEC) requires every timelike observer to measure a nonnegative local energy
density, whereas the strong energy condition (SEC) constrains the
trace-reversed tensor and supplies the matter term in timelike-geodesic
convergence. Finally, the dominant energy condition (DEC) supplements
nonnegative measured energy with the requirement that the associated energy
flux be causal and future directed. For the present diagonal source, these
conditions reduce to
\begin{align}
\mathrm{NEC}:&\quad
\rho+p_r\geq0,\qquad \rho+p_\perp\geq0,
\nonumber\\
\mathrm{WEC}:&\quad
\rho\geq0,\qquad \rho+p_r\geq0,\qquad
\rho+p_\perp\geq0,
\nonumber\\
\mathrm{SEC}:&\quad
\rho+p_r\geq0,\qquad \rho+p_\perp\geq0,\qquad
\rho+p_r+2p_\perp\geq0,
\nonumber\\
\mathrm{DEC}:&\quad
\rho\geq0,\qquad |p_r|\leq\rho,\qquad
|p_\perp|\leq\rho .
\label{eq:energy_conditions_type_i}
\end{align}

\subsection{Common consequences of the Casimir--memory source}

The sign of the density is fixed analytically before any complexity
prescription is imposed on the temporal geometry. Using the source profile
\eqref{eq:rho} and the dimensionless parameters
\eqref{eq:dimensionless_parameters}, one finds
\begin{equation}
8\pi r_0^2\rho(x)
=-\frac{a}{x^4}+\frac{e}{x^7}
=\frac{e-a x^3}{x^7}<0,
\qquad 1\leq x\leq x_\Sigma .
\label{eq:negative_density_interior}
\end{equation}
Indeed, \(\alpha>0\) and \(\eta\geq0\), as specified in
Eq.~\eqref{eq:rho}, imply \(a>0\) and \(e\geq0\). Moreover,
Eqs.~\eqref{eq:beta_throat} and \eqref{eq:flareout} give
\(a=e+\beta'(r_0)>e\). Hence \(a x^3\geq a>e\) for \(x\geq1\), and the
numerator in Eq.~\eqref{eq:negative_density_interior} is strictly negative.

This result is independent of the complexity closure because all the
solution families considered here share the same prescribed density and
shape function. The WEC therefore fails throughout the matter-supported
region for the exact vanishing-complexity, \(p\)-deformed, and
nonconstant locally vanishing-complexity lapse sectors. Since the DEC
requires, in particular, a nonnegative energy density, it also fails
throughout the interior for all these constructions.

A second common result follows directly at the throat. Equations
\eqref{eq:throat_density} and \eqref{eq:throat_radial_nec}, together with
Eq.~\eqref{eq:pr_einstein}, give
\(8\pi r_0^2p_r(r_0)=1\), while the hyperbolic flaring-out condition implies
\(\rho(r_0)+p_r(r_0)<0\). Thus the radial pressure at the throat is positive,
but the radial NEC nevertheless fails because the negative energy density
has the larger magnitude. This violation is therefore a geometrical
consequence of the hyperbolic flaring-out condition and does not depend on
which of the complexity prescriptions is subsequently used to determine the
redshift function.

These source- and throat-level statements constitute the conclusions that can
be transferred directly to the nonconstant locally vanishing-complexity
lapse branch without further information about its pressure profiles. The
remaining null and strong energy-condition combinations involve the temporal
geometry explicitly and must therefore be examined within a specified lapse
sector.

\subsection{Pressure-dependent conditions in the \(p\)-deformed family}

We now restrict the detailed analysis to the regular \(p\)-deformed family
of Eq.~\eqref{eq:pgeneral}. This family contains the exact
vanishing-complexity solution as the particular case \(p=3\) and allows the
role of a controlled residual complexity to be followed continuously.

For the finite-derivative branch of Eq.~\eqref{eq:pgeneral}, the throat
limit fixes
\(du/dx=[p-\beta'(r_0)]/[\beta'(r_0)-1]\) at \(x=1\).
Substitution into the angular Einstein equation gives
\begin{align}
8\pi r_0^2p_\perp(r_0)
&=\frac{p-1}{2},
\nonumber\\
8\pi r_0^2[\rho+p_\perp]_{r_0}
&=\frac{p-1-2\beta'(r_0)}{2},
\nonumber\\
8\pi r_0^2[\rho+p_r+2p_\perp]_{r_0}
&=p-\beta'(r_0).
\label{eq:throat_tangential_conditions}
\end{align}
Consequently, tangential NEC satisfaction at the throat would require
\begin{equation}
p\geq1+2\beta'(r_0).
\label{eq:tangential_nec_requirement}
\end{equation}
This condition, however, is incompatible with the existence of a
pressure-free Darmois surface in the finite-\(r\)-derivative branch. As
shown in Sec.~IV, every such matching configuration must satisfy
\begin{equation}
p<\beta'(r_0).
\label{eq:energy_matching_bound}
\end{equation}
Since the hyperbolic flaring-out condition requires \(\beta'(r_0)>1\),
one necessarily has
\begin{equation}
p<\beta'(r_0)<1+2\beta'(r_0).
\end{equation}
Therefore,
\begin{equation}
(\rho+p_\perp)_{r_0}<0
\label{eq:tangential_nec_matching_violation}
\end{equation}
for every regular pressure-free matching configuration in the
finite-\(r\)-derivative \(p\)-family. Thus the tangential NEC violation at
the throat is not merely a feature of the numerical examples; it follows
analytically from the simultaneous requirements of hyperbolic flare-out
and shell-free matching.

The same matching bound fixes the sign of the strong-energy trace
combination at the throat. From the last line of
Eq.~\eqref{eq:throat_tangential_conditions},
\begin{equation}
8\pi r_0^2[\rho+p_r+2p_\perp]_{r_0}
=p-\beta'(r_0)<0,
\end{equation}
for every pressure-free matching configuration in this branch. The
\(p=3\) solution is isotropic at the throat,
\(p_\perp(r_0)=p_r(r_0)\), but not necessarily away from it. Notice that
the SEC is already violated there independently through the radial null
inequality; the result above shows in addition that its trace combination
is necessarily negative throughout the matching-admissible
finite-\(r\)-derivative family.

The other physically distinguished point is the Darmois surface. Since
\(p_r(r_\Sigma)=0\) by Eq.~\eqref{eq:darmois_pressure}, the radial null
contraction on the interior side reduces to \(\rho(r_\Sigma^-)\), which is
negative by Eq.~\eqref{eq:negative_density_interior}. The radial NEC is
therefore violated both at the throat and immediately before the matching
surface throughout the regular \(p\)-deformed family. In the hyperbolic
vacuum, \(T_{\mu\nu}=0\), and all four energy conditions are saturated. The
interior limits of \(\rho\) and \(p_\perp\) need not vanish at \(r_\Sigma\):
the Darmois construction excludes a distributional surface layer but does
not require every bulk stress component to approach zero.

\begin{figure}[!h]
    \centering
    \includegraphics[width=0.96\textwidth]{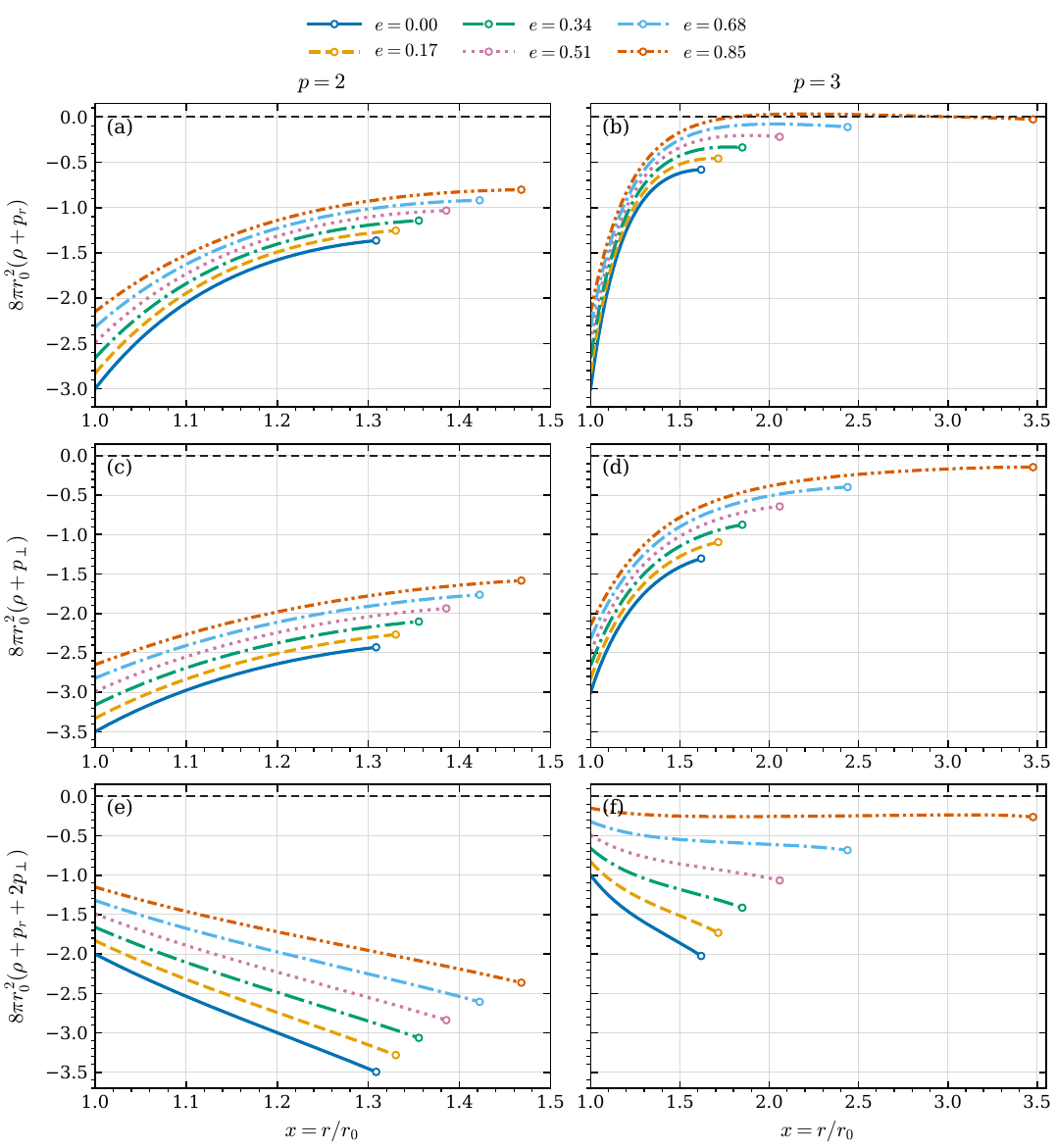}
    \caption{
    Dimensionless energy-condition combinations throughout the
    matter-supported region at fixed \(a=4\). The left and right columns
    correspond to \(p=2\) and \(p=3\), respectively. Panels (a) and (b)
    show \(8\pi r_0^2(\rho+p_r)\), panels (c) and (d) show
    \(8\pi r_0^2(\rho+p_\perp)\), and panels (e) and (f) show
    \(8\pi r_0^2(\rho+p_r+2p_\perp)\). The six curves correspond to
    \(e=0.00,\,0.17,\,0.34,\,0.51,\,0.68,\) and \(0.85\), while
    \(x=r/r_0\) is defined in Eq.~\eqref{eq:dimensionless_parameters}.
    Dashed horizontal lines mark saturation of the corresponding
    inequality. Open circles mark each curve's first \(p_r=0\) Darmois
    surface; no curve is extrapolated into the hyperbolic vacuum.}
    \label{fig:ec_profiles}
\end{figure}

For the profiles between these two boundaries, it is useful to eliminate
\(u''\) between Eqs.~\eqref{eq:pgeneral} and
\eqref{eq:pperp_einstein}. The resulting form,
\begin{equation}
8\pi p_\perp
=
\frac{2}{r}\left(\frac{\beta}{r}-1\right)\frac{u'}{u}
+
\frac{r\beta'-\beta+r_0[p-\beta'(r_0)]}{2r^3},
\label{eq:regular_tangential_pressure}
\end{equation}
is regular in the throat limit and is the expression used in the numerical
evaluation below. Figure~\ref{fig:ec_profiles} compares the regular \(p=2\)
and \(p=3\) solutions at fixed \(a=4\) for six values of the memory
parameter between \(e=0\) and \(e=0.85\).
Across the displayed family, increasing \(e\) shifts all three combinations
upward and moves the matching surface outward. This trend is consistent
with the reduction of the throat slope \(\beta'(r_0)=a-e\) at fixed \(a\),
together with the throat relations in
Eq.~\eqref{eq:throat_tangential_conditions}. The change in the material
extent is modest for \(p=2\), where \(x_\Sigma\) increases from
approximately \(1.309\) to \(1.468\), but is much stronger for \(p=3\),
where it increases from approximately \(1.619\) to \(3.477\).

For \(p=2\), panels (a), (c), and (e) remain below their saturation lines
for all six profiles and throughout every displayed material interval.
Thus, both independent null inequalities and the trace combination entering
the SEC are violated for these sampled configurations. The upward
displacement with increasing \(e\) weakens the magnitude of the violations
but does not change their signs before the Darmois surface.

The \(p=3\) family displays a more differentiated response. For
\(e\leq0.68\), all three combinations remain negative. At \(e=0.85\),
however, the radial null contraction in panel (b) crosses zero near
\(x\simeq1.805\), reaches a small positive maximum, and becomes negative
again near \(x\simeq3.058\), before terminating at
\(x_\Sigma\simeq3.477\). This finite positive interval represents local
satisfaction of the radial null inequality only. The tangential null
contraction in panel (d) and the trace combination in panel (f) remain
negative, so neither the complete NEC nor the SEC is restored. The WEC and
DEC also remain violated because the density is strictly negative by
Eq.~\eqref{eq:negative_density_interior}.

The numerical profiles therefore refine, rather than replace, the analytic
boundary results. The common Casimir--memory source fixes the negative
density and the associated WEC and DEC violations independently of the
lapse branch, while hyperbolic flare-out fixes the radial NEC violation at
the throat. For the matching-admissible finite-\(r\)-derivative
\(p\)-family, the bound \(p<\beta'(r_0)\) further implies analytically that
both the tangential null contraction and the strong-energy trace
combination are negative at the throat. These latter statements are
specific to that branch and are not being transferred to the nonconstant
locally vanishing-complexity solution of Sec.~V.

Within the \(p\)-deformed family sampled here, increasing the memory
contribution can nevertheless create an intermediate radial sector in
which the radial null inequality is locally satisfied for \(p=3\). As
shown in Fig.~\ref{fig:ec_profiles}, this local recovery does not extend
to the complete NEC or SEC, and the WEC and DEC remain violated because
the density is strictly negative.

\section{Conclusions}
\label{sec:conclusions}
\begingroup
\setlength{\parskip}{0pt}
\setlength{\parindent}{1.5em}

We have developed a unified construction of traversable hyperbolic wormholes sustained by a Casimir vacuum supplemented by a phenomenological gravitational-memory correction. In the present matter-first approach, the prescribed Casimir--memory density determines the radial geometry analytically through the Einstein equations, whereas the temporal sector is selected by complexity-based conditions. This makes it possible to compare different lapse sectors without changing the underlying matter source or shape function, so that the resulting configurations form a connected family rather than a collection of unrelated wormhole models.

The memory contribution changes the throat derivative, restricts the local flaring-out domain, modifies the redshift profile, and shifts the location and existence of admissible matching surfaces. Its coefficient is motivated by the residual Casimir-energy shift produced by a weak transient gravitational perturbation, but its implementation as a radial correction remains phenomenological. Accordingly, the present results should be interpreted as an investigation of how a memory imprint of this type would affect a hyperbolic wormhole, rather than as a microscopic derivation of the corresponding radial source from the gravitational pulse itself.

For exact vanishing complexity, the redshift function is fixed by a second-order equation whose finite-$r$-derivative throat condition determines the remaining derivative once the temporal normalization is chosen. Regular configurations exist over finite regions of parameter space and terminate at the first regular zero of the radial pressure. At that pressure-free surface, the interior can be matched smoothly to the hyperbolic Schwarzschild vacuum through the Darmois conditions, with no surface layer required. The absence of a thin shell is therefore not imposed by construction; it emerges only for those source and complexity parameters for which a regular pressure-free junction is reached before the loss of the admissible static interior.

The one-parameter complexity deformation provides a controlled departure from exact vanishing complexity. The deformation parameter directly measures the amplitude of a residual inverse-cubic contribution to the complexity factor while leaving the prescribed density and radial geometry unchanged. Varying this parameter changes the lapse and the matching domain. A central analytical result is that a pressure-free Darmois surface in the finite-$r$-derivative branch can exist only when the deformation parameter remains below the throat derivative. The constant-redshift, zero-tidal-force configuration occurs when these two quantities coincide and therefore lies precisely at the limiting value beyond which this finite-$r$-derivative branch cannot reach a pressure-free matching surface.

The decomposition of the full complexity factor into local and throat contributions reveals that local vanishing complexity is not an independent third closure, but a distinguished sector of the same deformed construction, selected when the deformation parameter equals the throat derivative. At this value, the finite-$r$-derivative solution reduces to the constant-redshift branch, whereas a second independent solution yields a nonconstant lapse. Although its derivative with respect to the areal radius diverges at the throat, this second branch is regular in proper radial distance and has finite curvature there. Its lapse is obtained semianalytically by a single quadrature. The general Darmois matching fixes the remaining lapse integration freedom together with the relative normalization between the interior and exterior time coordinates. The additional unit normalization used in the numerical examples is therefore only a particular subclass of this nonconstant branch. For some source parameters, the corresponding matching condition admits more than one admissible radius, and the representative solutions displayed here use the outermost one.

The energy-condition analysis clarifies which features are intrinsic to the common Casimir--memory source and hyperbolic throat and which depend on the lapse branch. The flaring-out requirement enforces a negative energy density throughout the matter-supported region, so the weak and dominant energy conditions fail independently of the temporal sector. The radial null energy condition is necessarily violated at the throat even though the radial pressure there is positive; this is a direct consequence of hyperbolic flaring-out rather than of a negative radial pressure. For the matching-admissible finite-$r$-derivative deformed family, the analytical matching bound further implies that both the tangential null contraction and the strong-energy trace combination are necessarily negative at the throat. These violations are therefore not merely features of the numerical examples. Away from the throat, the radial null combination may nevertheless become positive over an intermediate interval for some configurations. The nonconstant locally vanishing-complexity branch shares the same negative density, and hence the same weak- and dominant-energy-condition violations, while its pressure-dependent null and strong conditions require a separate detailed analysis.

After the Darmois matching, the memory parameter does not survive as an independent exterior charge. Its effect is instead encoded indirectly in the interior radial geometry, the lapse, the pressure-free junction data, and the effective Schwarzschild mass. The resulting picture links the Casimir--memory source, hyperbolic flaring-out, complexity, and shell-free vacuum matching within a single framework. The organizing structure is therefore more constrained than a simple progression among three independent complexity prescriptions: exact vanishing complexity belongs to the deformed family at a particular parameter value, while local vanishing complexity selects another distinguished value and exposes a second, proper-distance-regular lapse branch.

Natural extensions include the stability analysis of the admissible branches, dynamical generalizations of the present static configurations, and a more microscopic treatment of the memory-corrected vacuum source. The nonconstant locally vanishing-complexity solution also raises the question of how globally asymmetric but smooth two-sided extensions should be characterized dynamically. It would further be useful to investigate whether the dependence on the memory sector that remains encoded in the junction data admits a meaningful quasi-local characterization.

\endgroup

\section*{Acknowledgments}
CRM would like to thank Conselho Nacional de Desenvolvimento Cient\'{i}fico e Tecnol\'ogico (CNPq) for the partial financial support, through grant 301122/2025-3. FBL is funded by Fundação Cearense de Apoio ao Desenvolvimento Científico e Tecnológico (FUNCAP) and by  Conselho Nacional de Desenvolvimento Científico e Tecnológico (CNPq), grant number 305947/2024-9.

\bibliographystyle{apsrev4-1}
\bibliography{ref_short.bib}

\end{document}